\documentclass[fleqn,usenatbib]{mnras}

\usepackage{newtxtext,newtxmath}

\usepackage[T1]{fontenc}
\usepackage{threeparttable}
\usepackage{threeparttablex}
\usepackage{adjustbox}
\usepackage{float}
\usepackage{comment}
\usepackage{ulem}
\DeclareRobustCommand{\VAN}[3]{#2}
\let\VANthebibliography\thebibliography
\def\thebibliography{\DeclareRobustCommand{\VAN}[3]{##3}\VANthebibliography}
\DeclareUnicodeCharacter{2212}{-}

\usepackage{graphicx}	
\usepackage{amsmath}	

\title[SN~2020aze]{Early interaction signatures and an extended plateau phase in Type II SN~2020aze}
\author[B. Ailawadhi et al.]{B. Ailawadhi,$^{1,2,3}$\thanks{E-mail: bhavya@prl.res.in; K. A. Bostroem is an LSST-DA Catalyst Fellow}
R. Dastidar,$^{4}$
K. Misra,$^{1}$
S. Valenti,$^{5}$
D. J. Sand,$^{6}$ 
J. E.  Andrews,$^{7}$
J. P. Anderson,$^{8}$
\newauthor
K. A. Bostroem,$^{6}$
P. J. Brown,$^{9}$
R. Cartier, $^{10}$
T. W. Chen, $^{11}$
Y. Dong, $^{5}$
N. Dukiya,$^{1,12}$
E. Padilla Gonzalez,$^{13}$
\newauthor
M. Gromadzki,$^{14}$
J. Haislip,$^{15}$
D. Hiramatsu,$^{16,17}$
D. A. Howell,$^{18,19}$
C. Inserra,$^{20}$
D. Janzen,$^{21}$
S. W. Jha,$^{22}$
\newauthor
V. Kouprianov,$^{15}$
C. McCully,$^{15}$
T. E. Müller-Bravo, $^{23,24}$
C. Pellegrino,$^{25}$
G. Pignata,$^{26}$ 
D. E. Reichart,$^{15}$
\newauthor
J. Sollerman,$^{27}$
D. R. Young,$^{28}$
L. Yadav$^{2}$
\\
$^{1}$ Aryabhatta Research Institute of Observational Sciences, Manora Peak, Nainital 263 001, India\\
$^{2}$ Department of Physics, Deen Dayal Upadhyaya Gorakhpur University, Gorakhpur-273009, India\\
$^{3}$ Astronomy \& Astrophysics Division, Physical Research Laboratory, Ahmedabad-380009, India\\
$^{4}$ Istituto Nazionale di Astrofisica, Osservatorio Astronomico di Brera, via E. Bianchi 46, 23807 Merate (LC), Italy\\
$^{5}$ Department of Physics and Astronomy, University of California, 1 Shields Avenue, Davis, CA 95616-5270, USA\\
$^{6}$ Steward Observatory, University of Arizona, 933 North Cherry Avenue, Tucson, AZ 85721-0065, USA\\
$^{7}$ Gemini Observatory, 670 N. Aohoku Place, Hilo, HI 96720, USA\\
$^{8}$ European Southern Observatory, Alonso de C\'ordova 3107, Casilla 19, Santiago, Chile\\
$^{9}$ George P. and Cynthia Woods Mitchell Institute for Fundamental Physics \& Astronomy, Texas A\&M University, 4242 TAMU, College Station, TX 77843, USA\\
$^{10}$ Centro de Astronom´ıa (CITEVA), Universidad de Antofagasta, Avenida Angamos 601, Antofagasta, Chile\\
$^{11}$ Graduate Institute of Astronomy, National Central University, 300 Jhongda Road, 32001 Jhongli, Taiwan\\
$^{12}$ Department of Applied Physics, Mahatma Jyotiba Phule Rohilkhand University, Bareilly, 243006, India\\
$^{13}$ Space Telescope Science Institute, 3700 San Martin Drive, Baltimore, MD 21218, USA\\
$^{14}$ Astronomical Observatory, University of Warsaw, Al. Ujazdowskie 4, 00-478 Warszawa, Poland\\
$^{15}$ Department of Physics and Astronomy, University of North Carolina, 120 East Cameron Avenue, Chapel Hill, NC 27599, USA\\
$^{16}$ Center for Astrophysics | Harvard \& Smithsonian, 60 Garden Street, Cambridge, MA 02138-1516, USA\\
$^{17}$ The NSF AI Institute for Artificial Intelligence and Fundamental Interactions, USA\\
$^{18}$ Department of Physics, University of California, Santa Barbara, CA 93106-9530, USA\\
$^{19}$ Las Cumbres Observatory, 6740 Cortona Drive, Suite 102, Goleta, CA 93117-5575, USA\\
$^{20}$ Cardiff Hub for Astrophysics Research and Technology, School of Physics \& Astronomy, Cardiff University, Queens Buildings, The Parade, Cardiff, CF24 3AA, UK\\
$^{21}$ Department of Physics \& Engineering Physics, University of Saskatchewan, 116 Science Place, Saskatoon, SK S7N 5E2, Canada\\
$^{22}$ Rutgers University, Department of Physics and Astronomy, 136 Frelinghuysen Road Piscataway, NJ 08854, USA\\
$^{23}$ School of Physics, Trinity College Dublin, The University of Dublin, Dublin 2, Ireland\\
$^{24}$ Instituto de Ciencias Exactas y Naturales (ICEN), Universidad Arturo Prat, Chile\\
$^{25}$ NASA/Goddard Space Flight Center, Greenbelt, MD, USA\\
$^{26}$ Instituto de Alta Investigaci\'on, Universidad de Tarapac\'a, Casilla 7D, Arica, Chile\\
$^{27}$ Oskar Klein Centre, Department of Astronomy, Stockholm University, AlbaNova, SE-106 91 Stockholm, Sweden\\
$^{28}$ Astrophysics Research Centre, School of Mathematics and Physics, Queen’s University Belfast, Belfast BT7 1NN, UK
}

\date{Accepted XXX. Received YYY; in original form ZZZ}

\pubyear{2025}

\begin{document}
\label{firstpage}

\pagerange{\pageref{firstpage}--\pageref{lastpage}}

\maketitle


\begin{abstract}
\label{abstract}
\vspace{-0.2mm}
We present a photometric and spectroscopic analysis of the fast-declining Type II SN~2020aze, observed in optical bands from 2.2 to 137.4 days post-explosion. The $V$-band light curve reaches a peak absolute magnitude of $-$16.97$\pm$0.20 mag by 15 days, followed by a recombination phase with a decline rate of $2.04\pm0.13$ mag (100 day)$^{-1}$, lasting $\sim$120 days. Early spectra ($<$6.0 day) exhibits a transient weak narrow emission line at 4687 \AA{} and a bump or ledge feature spanning 4400–4800 \AA{}, attributed to narrow and broad blue-shifted \ion{He}{ii} $\lambda$4686, indicating interaction between the rapidly expanding ejecta and dense circumstellar material (CSM). Spectral comparison with literature models suggests a red supergiant progenitor with a weak wind and a mass-loss rate of $\sim$10$^{-3}$ M$_\odot$ yr$^{-1}$. Semi-analytical light curve modelling yields an initial radius of 1100 R$_{\odot}$, an ejecta mass of 12 M$_\odot$, a total explosion energy of 1.5$\times$10$^{51}$ erg, and a progenitor mass of approximately 14 M$_\odot$. The combination of a steep luminosity decline, early interaction signatures, and an unusually extended photospheric phase highlights the complex interplay between pre-SN mass loss, CSM interaction, and progenitor properties, positioning SN~2020aze as an important case for understanding the diversity of Type II SNe.

\end{abstract}


\newpage
\section{Introduction}
\label{introduction}

Core-collapse supernovae (CCSNe) are catastrophic events marking the end stages of stars with masses $\gtrsim$ 8\(M_\odot\) \citep{Woosley2002, Smartt2009_progenitors}. These occur when the gravitational forces overcomes the thermal pressure at the end of nuclear burning. Type II SNe fall under the CCSNe category and are characterised by hydrogen features in their spectra \citep{filippenko1997}. They are further divided into subtypes: Type IIL, IIP, IIn, and IIb, based on their light curve and spectroscopic characteristics. Type IIL SNe exhibit a linear decline during the hydrogen recombination phase, which is also referred to as the photospheric phase, of their light curves, while Type IIP SNe display a plateau lasting 80 to 120 days during this phase \citep{barbon}. Type IIn SNe show narrow spectral features arising from the interaction between the expanding ejecta and the circumstellar medium \citep[CSM;][]{Schlegel1990, Chugai2001}. Type IIb SNe initially exhibit prominent hydrogen features in their spectra, which gradually fade over time, while helium lines become increasingly dominant, reflecting a progenitor that retained only a thin hydrogen envelope at the time of explosion \citep{filippenko1997}. 

The hydrogen-rich Type II SNe exhibit a wide range of properties, encompassing both Type IIP and Type IIL events, which together form a continuum of features \citep{anderson2014, Sanders, Galbany}. Traditionally, the distinctions between Type IIP and IIL SNe were based on their decline rates, with Type IIL SNe defined as having a decline rate greater than 1 mag (100 days)$^{-1}$ during the recombination phase \citep{Faran2014}. Type IIL SNe are generally brighter than Type IIP SNe \citep{anderson2014, Faran2014, dejaeger2019}, a difference attributed to the presence of a substantial hydrogen envelope in Type IIP progenitors, which prolongs the release of shock-deposited energy. In contrast, Type IIL progenitors are thought to have lower hydrogen content (1–2 M$_{\odot}$, \citealt{Blinnikov, hiramatsu2021_2018zd}), leading to faster declines in luminosity, shorter photospheric phases, and higher peak luminosities. Lower hydrogen content can also result in higher ejecta velocities \citep{gutirrez2017a}, although interactions between ejecta and CSM in some cases produce features more reminiscent of Type IIP SNe, such as extended photospheric phases and lower velocities \citep{Hillier2019}. With an increasing number of Type II SNe observed, it has become evident that the distinctions between Type IIP and IIL SNe are not always clear-cut, as multiple factors, such as hydrogen envelope mass, CSM interaction and ejecta density, contribute to their observed diversity \citep{anderson2014, valenti16}. Given this blurred domain, we collectively refer to both Type IIP and IIL as Type II SNe for the rest of this paper.

In recent years, advances in observational facilities have made it possible to observe numerous young Type II SNe, some of which display signs of CSM interaction, evidenced by narrow high ionisation emission lines of \ion{He}{ii}, \ion{C}{iii}, \ion{C}{iv}, \ion{N}{iv}, \ion{N}{v}, and \ion{O}{iv}, as well as narrow \ion{H}{i} lines in spectra taken within the first week after explosion (e.g. \citealt{yaron2017, Boian2019, Tartaglia2017ahn, Bruch2021, azalee_2023ixf, Jacobson2024, shrestha2024}). These highly ionised features are valuable diagnostics of the density, velocity and composition of CSM ejected in the years prior to explosion \citep{Boian2019}. \cite{Khazov2016} found that around 20\% of Type II SNe discovered within 5 days of explosion show these ``flash-ionisation'' features. Moreover, \cite{Bruch2023} reported that more than 36\% of early-stage SNe observed within 2 days post-explosion showed these spectral lines. These features are typically observed in Type IIn SNe for a much longer duration, where shock breakout occurs within the surrounding CSM. This process continues as long as the optical depth of the CSM lying above the shock front exceeds {\textit {c/v}}, where {\textit v} corresponds to the shock expansion velocity. In Type II SNe, the interaction signatures last from a few hours to a few days until the ejecta have swept up the CSM. The full width at half maximum (FWHM) of these lines is estimated to be around 10$^2$ km s$^{-1}$ (e.g. SN 2023ixf; \citealt{azalee_2023ixf}, SN 2024ggi; \citealt{shrestha2024}).  

This study presents photometric and spectroscopic analysis of SN~2020aze, a fast-declining Type II SN with early CSM interaction and an unusually extended photospheric phase. Its hybrid characteristics offer a rare opportunity to examine how progenitor mass loss, CSM interaction, and explosion physics drive the diversity among Type II SNe. The structure of the paper is as follows. Section \ref{sn2020aze_properties} details the detection, SN properties and the data acquisition and reduction. Section \ref{lightcurve_analysis} covers the light curve analysis, while Section \ref{LC modelling} presents the light curve modelling. In Section \ref{spectra}, we discuss the spectroscopic analysis, highlighting the early flash features observed in the early spectra. Section \ref{Parallels of SN 2020aze} explores potential SN classes that SN~2020aze may belong to. Finally, the main conclusions of this study are summarised in Section \ref{discussion}.

\section{SN~2020\texorpdfstring{\lowercase{aze}}{aze}}
\label{sn2020aze_properties}
\subsection{Basic information and explosion epoch}
\begin{figure}
\includegraphics[width=\columnwidth]{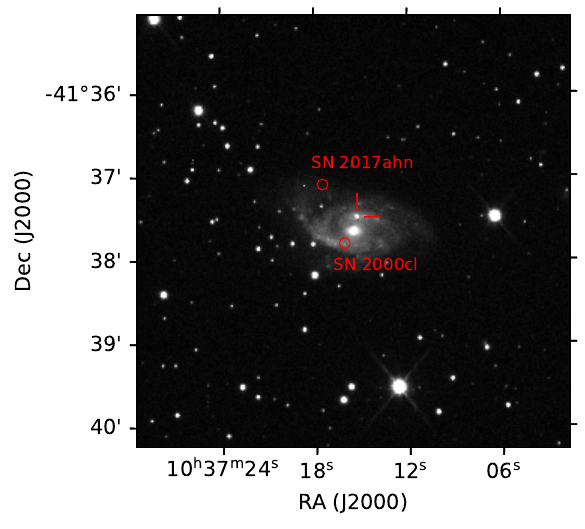}
    \caption{A 120-second $r$-band image of the host galaxy NGC~3318, taken with the 1-m LCO telescope on MJD 58876.9, showing the positions of SN~2020aze along with two previous SNe~2000cl and 2017ahn. It is along the north-East direction.}
    \label{Fieldimage}
\end{figure}

\begin{figure}
\includegraphics[width=\columnwidth]{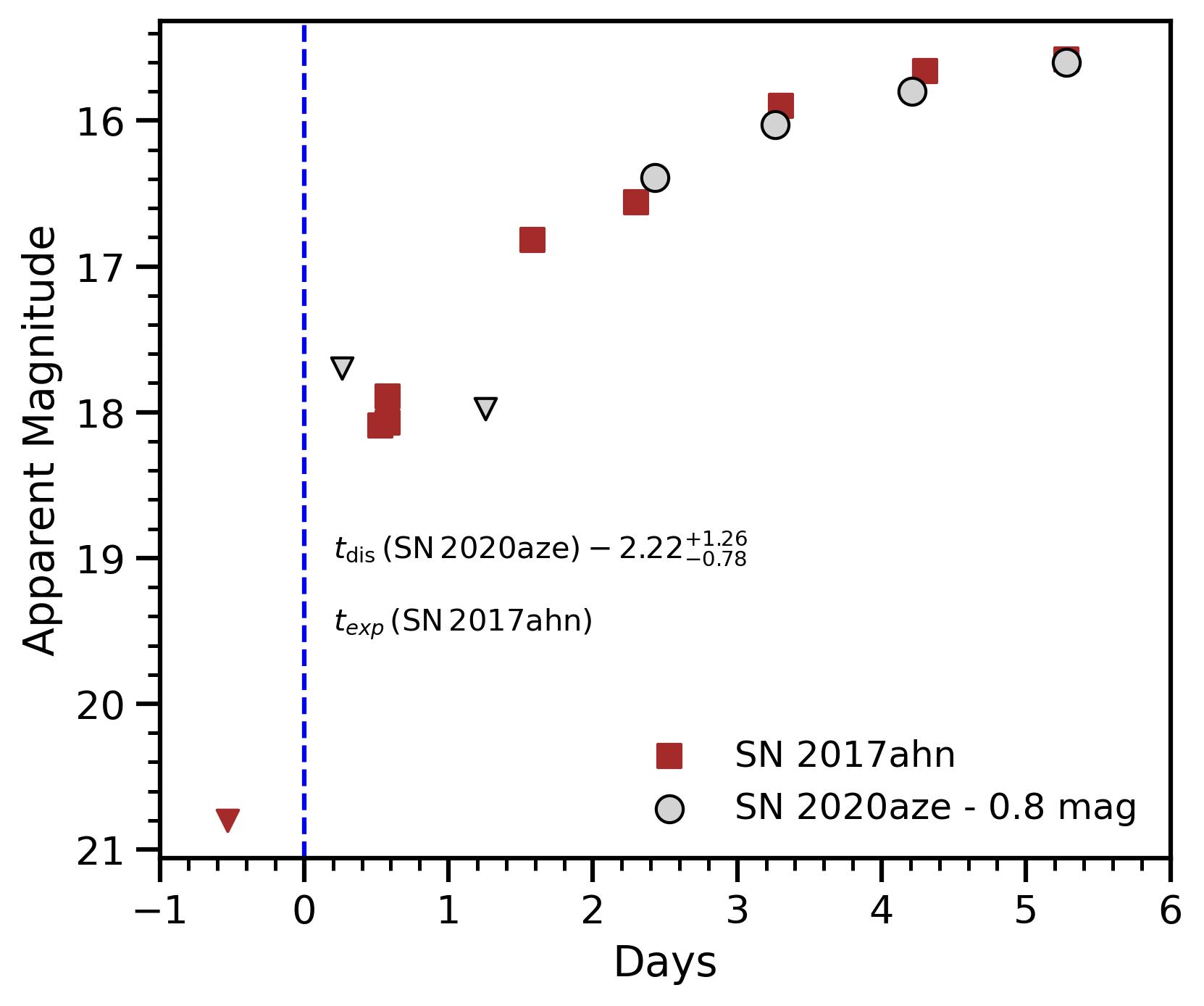}
    \caption{DLT40 \texttt{Clear}-band light curve of SN~2017ahn is matched to that of SN~2020aze during the rising phase. The x-axis denotes the days since the explosion for SN~2017ahn. The best fit to the rising light curve (up to 6 days post-discovery) is used to constrain the explosion epoch of SN~2020aze more precisely as compared to the shallow DLT40 non-detection limits prior to discovery (inverted triangle).}
    \label{explosionepoch}
\end{figure}

SN~2020aze (also known as DLT20d) was discovered by the Distance Less Than 40 (DLT40) survey group \citep{Tartaglia2018} at $\mathrm{R.A.} = 10^\mathrm{h}37^\mathrm{m}15.27^\mathrm{s}$, $\mathrm{Dec} = -41\degr37^{\prime}28\farcs92$ (J2000.0) on UT 26-01-2020 12:43:12.00 (MJD=58874.53) with the PROMPT5 0.4 m telescope \citep{Valenti_disc}. The SN exploded $3\farcs60$ W and $10\farcs28$ N from the centre of face-on spiral galaxy NGC~3318. Previously, two other Type II SNe~2000cl \citep{Chassagne} and 2017ahn \citep{Tartaglia2017ahn} were also discovered in NGC~3318. The location of SN~2020aze, along with the locations of SNe~2000cl and 2017ahn, are shown in Figure~\ref{Fieldimage}.
The last DLT40 non-detection of SN~2020aze was on MJD~58873.54, at a limiting magnitude of 18.77~ABmag in the Clear filter, approximately one day prior to the first detection (17.12~mag in the Clear band on MJD~58874.53). While this non-detection provides a coarse constraint, its relatively shallow limiting magnitude suggests that the explosion may have occurred earlier than the midpoint of these two epochs. To derive a more precise explosion date, we compared the early-time light curve of SN~2020aze to that of SN~2017ahn, a similar event discovered in the same host galaxy with much deeper non-detection limits. We performed template fitting to match the rising part of the light curve of SN~2020aze (up to 6 days from discovery) to the well-constrained rise of SN~2017ahn. The best-fit match requires a shift of $−$2.22~days relative to the discovery of SN~2020aze (as shown in Figure~\ref{explosionepoch}). Given the discovery date of MJD~58874.53, this template-matching method yields an estimated explosion epoch of 58872.31~MJD, which we adopt for our analysis.

The SN was classified as a Type II SN \citep{Irani_classi} based on the spectrum obtained with EFOSC2 mounted on the ESO New Technology Telescope (NTT) at La Silla on MJD 58881.2 (7 days after the discovery). Table~\ref{tab:sn2020aze} presents the basic information of SN~2020aze and its host galaxy.

\begin{table}
    \centering
    \caption{Basic information of SN~2020aze and the host galaxy NGC~3318.}
    \begin{tabular}{ll} 
		\hline
            \multicolumn{2}{c}{\textbf{SN 2020aze}} \\ \hline
            Discovery Date (MJD) & 58874.53\\
            Explosion Date$^{\dagger}$ (MJD) & 58872.31$^{+1.26}_{-0.78}$\\
            SN type &  II\\
            $\mathrm{R.A.}$ (J2000) & $10^\mathrm{h}37^\mathrm{m}15^\mathrm{s}.27$\\$\mathrm{Dec}$ (J2000) & $-41\degr37^{\prime}28\farcs92$\\
            Discovery Magnitude & 17.12 mag (Clear)\\
            {\it E(B-V)}$^{\dagger}$ (host + MW) & 0.20$\pm$0.02 mag\\
            \hline \multicolumn{2}{c}{\textbf{NGC 3318}$^{\ddagger}$} \\ \hline
		Galaxy Type & SBbc D\\
		Major Diameter & 2$^\prime$.52\\
		Minor Diameter & 1$^\prime$.36\\
            Redshift & 0.00926\\
		Distance & 36.7$\pm$2.6 Mpc\\
		Helio. Velocity & 1566.12$\pm$2.10 km s$^{-1}$\\
		
		\hline
    \end{tabular}
    \newline
	\noindent
$^{\dagger}${This work},
$^\ddagger${From NED}\\
	\noindent
	\label{tab:sn2020aze}
\end{table}

\subsection{Data acquisition and reduction}
\label{Data}

The DLT40 \texttt{Clear} band observations of SN~2020aze span from 2.22 to 119.14 days since the explosion. The multi-band follow-up observations in {\it UBVg}$^\prime${\it r}$^\prime${\it i}$^\prime$ were conducted from 2.6 to 137.4 days since explosion at 35 epochs under the Global Supernova Project (GSP) key program with the Las Cumbres Observatory (LCO) network \citep{Brown2013}. The LCO data are reduced using the \texttt{lcogtsnpipe}\footnote{\url{https://github.com/svalenti/lcogtsnpipe}} pipeline described in \citet{valenti16}. SN~2020aze was located close to the host galaxy nucleus; hence, a careful template image subtraction to estimate the SN magnitude free from host galaxy contamination was necessary. The SN field images observed on 03-02-2025 were used as templates for subtraction. High Order Transform of PSF ANd Template Subtraction (HOTPANTS\footnote{\url{https://github.com/acbecker/hotpants}}) was used to perform the image subtraction. The instrumental magnitudes of the SN in the {\it g}$^\prime${\it r}$^\prime${\it i}$^\prime$ bands were calibrated using reference stars from the AAVSO Photometric All-Sky Survey (APASS\footnote{\url{https://www.aavso.org/apass}}). For the {\it UBV}-band, calibration was performed using Landolt standard fields \citep{Landolt1983, Landolt1992} observed on the same night as the SN. The \texttt{Clear} band SN photometry from DLT40 and the {\it UBVg}$^\prime${\it r}$^\prime${\it i}$^\prime$ photometry from the LCO network are provided in Tables~\ref{dlt_phot} and \ref{optical_phot}, respectively. We have also used the \texttt{ATLAS} \citep{Tonry2018, Smith2020} forced photometry \citep{Shingles2021} \footnote{\url{https://fallingstar-data.com/forcedphot/}} (Table~\ref{atlas_phot}) in the \texttt{o} band (corresponding to \texttt{r+i} filter) in this work.

The UV photometry of SN 2020aze was obtained using the Ultra-Violet/Optical Telescope (UVOT) aboard the Neil Gehrels Swift Observatory (hereafter Swift; \citealt{Gehrels2004_swift}) between 3.6 and 10.6 days after explosion, covering five epochs across all UVOT filters. The optical UVOT data were discarded because the high background from the host galaxy made the non-linearity correction unreliable. Since the SN lies within its host galaxy, template subtraction was performed using pre-explosion images from 2019. Aperture photometry was carried out with a 5\arcsec aperture, following the procedures described in \citealt{Brown2009_UVOT} and \citealt{Poole2008}. The measured count rates were converted to AB magnitudes using the UVOT photometric zero points \citep{Poole2008, Breeveld2011}. The resulting count rates were transformed into AB magnitudes using the calibrated UVOT zero points reported by \citet{Poole2008} and \citet{Breeveld2011} (Table~\ref{UVOT_phot}).

The first spectrum, at 2.6 day post-explosion, was acquired with the Robert Stobie Spectrograph (RSS) mounted on the South African Large Telescope (SALT) with the PG0900 grating (resolution $\sim$5.6 \AA). The next spectroscopic observation, obtained at 3.0 day post-explosion, was carried out with the Gemini Multi-Object Spectrograph (GMOS) on the Gemini South (GS) Telescope \citep{Hook2004}. Subsequent spectroscopic observations up to $\sim$ 322 days since the explosion were taken with several low to medium resolution spectrographs mounted on the 2m LCO telescopes and 3.58 m NTT. The SN was also observed under the ePESSTO+ collaboration \citep{smartt2015_pessto}, using the EFOSC2 instrument on the NTT. One epoch spectrum was also obtained with the Goodman High Throughput Spectrograph (GHTS; \citealt{Clemens_2004}) on the Southern Astrophysical Research Telescope (SOAR). Additionally, the final spectrum at 322.0 day post-explosion was retrieved from the ESO public archive\footnote{\url{https://archive.eso.org/scienceportal/home}}. The log of spectroscopic observations is given in Table~\ref{spec_log}. The extraction of 1D-wavelength and flux calibrated spectra for the LCO data is done using the \texttt{floydsspec} pipeline\footnote{\url{https://github.com/svalenti/FLOYDSpipeline}} \citep{Valenti2014}. The EFOSC2 spectra were reduced using the PESSTO pipeline\footnote{\url{https://github.com/svalenti/pessto}} \citep{smartt2015_pessto} and the SALT spectrum was reduced using a pipeline based on \texttt{PySALT} package developed by \cite{crawford2010_pysalt}. The GMOS spectrum was reduced using the DRAGONS package \citep{Labrie2023} within the GMOS long-slit pipeline. The SOAR spectrum was reduced using the Goodman pipeline\footnote{\url{https://soardocs.readthedocs.io/projects/goodman-pipeline/en/latest/}}. To accommodate slit loss corrections, all spectra are scaled to match the photometric flux using the \texttt{lightcurve-fitting}\footnote{\url{https://github.com/griffin-h/lightcurve_fitting}} module \citep{lk_fitting_2022}. Finally, the spectra are corrected for the heliocentric redshift of the host galaxy.

\subsection{Distance and reddening}
\begin{figure}
\includegraphics[width=\columnwidth]{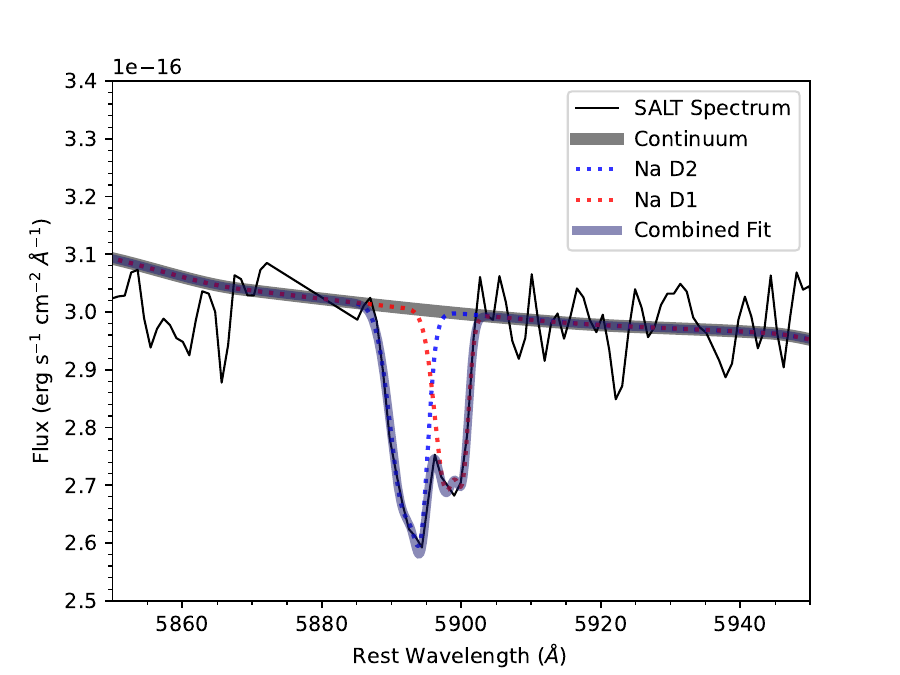}
\caption{Multi-component Gaussian fits to the distinct D2 and D1 features of the \ion{Na}{I} profile in the 2.6 day spectrum of SN~2020aze.}
\label{NaID}
\end{figure}

The luminosity distance, corresponding to a host galaxy redshift ($z$=0.009255) and corrected for the effects of Virgo Cluster, Great Attractor (GA), and Shapley superclusters, is estimated to be 36.7$\pm$2.6 Mpc\footnote{\url{https://ned.ipac.caltech.edu/byname?objname=NGC+3318&hconst=73&omegam=0.27&omegav=0.73&wmap=1&corr_z=4}} (assuming H$_0 = 73$ km s$^{-1}$ Mpc$^{-1}$, $\Omega_{\mathrm{matter}}=0.27$, $\Omega_{\mathrm{vacuum}}=0.73$) and is used throughout the paper. The Galactic reddening along the line-of-sight of SN~2020aze is $E(B-V)_{\rm MW}$ = 0.0667$\pm$0.0005 mag \citep{schlafly}. To estimate the host galaxy extinction, we use $A_{\mathrm{V}} = 0.78 \, (\pm 0.15). \, \mathrm{EW}_{\mathrm{Na\,ID}}$ \citep{Stritzinger2018}, which relates the equivalent width (EW) of the \ion{Na}{I} doublet ($\lambda\lambda$5890, 5896 ) to $E(B-V)$. Using the SALT spectrum with a resolution of R=1100, obtained 2.6 days after the explosion, we fit multi-component Gaussians to the distinctly visible D2 and D1 components of \ion{Na}{i} D (as shown in Figure~\ref{NaID}), yielding EWs of $0.77 \pm 0.08$ \AA{} and $0.54 \pm 0.05$ \AA, respectively. Applying the extinction law from \citet{Cardelli} with $R_V = 3.1$, the estimated $E(B-V)$ values are $0.19 \pm 0.04$ mag and $0.13 \pm 0.03$ mag. We multiply the error-weighted average value $0.15 \pm 0.02$ mag by 0.86 to be consistent with the recalibration of the Milky Way extinction by \cite{schlafly}, and obtain $E(B-V)_{\rm host}$ = $0.13 \pm 0.02$ mag. Thus, the total extinction from the Milky Way and the host galaxy is $E(B-V)_{\rm (MW+host)}$ = $0.20 \pm 0.02$ mag, which is used throughout this work.

\section{Light curve analysis}
\label{lightcurve_analysis}

\begin{figure}
\includegraphics[width=\columnwidth]{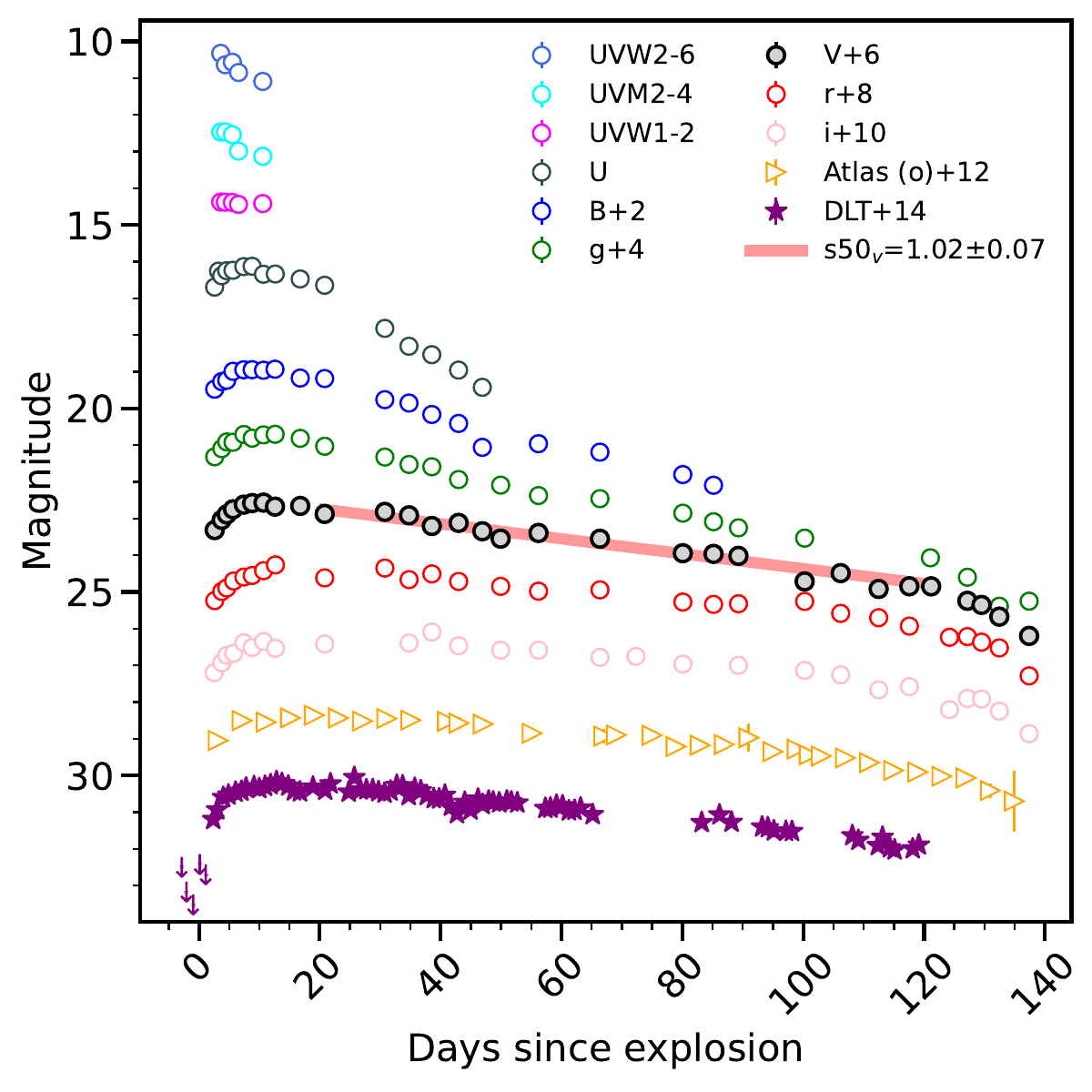}
    \caption{The multi-band light curves of SN~2020aze from 2.2 to 137.4 days since the explosion are shown. The red solid line marks the decline of the $V$-band light curve during the plateau phase (s$_2$), while the downward arrows indicate non-detections reported by the DLT40 survey.}
    \label{LC}
\end{figure}

The light curves of SN~2020aze, observed from 2.2 to 137.4 day since the explosion in various spectral bands, are shown in Figure~\ref{LC}. The slope (s50$_V$), shown by the red broad line in the figure, represents the decline rate of the V-band light curve in magnitudes per 50 days, measured over the 22 to 122 days post-explosion interval. The light curves show a rise to maximum brightness, followed by a linearly declining photospheric phase, and a subsequent drop marking the end of the photospheric phase. There are limited observations during the fall and no observations during the radioactive tail phase of SN~2020aze as it went behind the Sun. The peak time and magnitude in the various bands are estimated by fitting a polynomial to the observed magnitudes, and the values are provided in Table~\ref{peak_mag}. The peak in the bluer bands occurs much earlier than in the redder bands. The post-peak decline rates in the {\it U}, {\it B}, {\it g}$^\prime$, {\it V}, {\it r}$^\prime$, {\it i}$^\prime$ bands are also shown in Table~\ref{peak_mag}. It is evident that the redder bands display a flatter evolution compared to the light curves in the bluer bands. 

\begin{figure}
\includegraphics[width=\columnwidth]{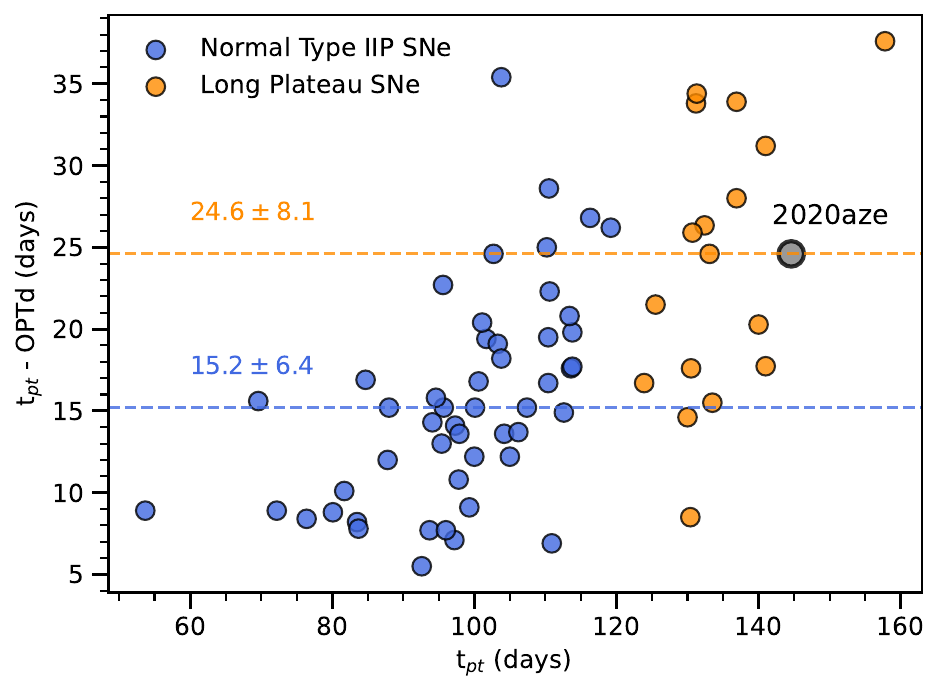}
    \caption{The scatter plot of $t_{\rm pt}$ − OPTd vs. $t_{\rm pt}$ for Type II SNe is shown. SNe with plateau durations shorter than 120 days are marked in blue, while those with durations longer than 120 days are marked in orange, and SN 2020aze is highlighted in grey. The horizontal lines represent the median value of $t_{\rm pt}$ − OPTd, and errors correspond to 1$\sigma$ of the distribution.}
    \label{tpt_optd}
\end{figure}

The duration of the optically thick phase (OPTd; \citealt{anderson2014}) in the $V$-band light curve of SN~2020aze, measured from the explosion epoch to the point where the linear decline ends, is approximately 120 days. Another commonly used indicator of the plateau length is $t_\mathrm{pt}$, defined as the midpoint of the transition from the plateau to the radioactive tail phase \citep{Olivares2010}, and thus extends beyond OPTd. Due to the limited observations during the decline phase of SN~2020aze, a direct estimate of $t_\mathrm{pt}$ from light curve fitting is not feasible. However, OPTd can still be measured. To estimate $t_\mathrm{pt}$, we used a sample of SNe II from \citet{anderson2014} and analysed the difference between $t_\mathrm{pt}$ and OPTd as a function of $t_\mathrm{pt}$. The median differences and corresponding 1$\sigma$ standard deviations for normal ($t_\mathrm{pt} < 120$ days) and long-plateau ($t_\mathrm{pt} > 120$ days) SNe are indicated in Figure~\ref{tpt_optd}. For long-plateau SNe IIP, $t_\mathrm{pt}$ typically exceeds OPTd by $\sim$24.6$\pm$8.1 days. Applying this relation to SN~2020aze gives an estimated plateau length of approximately 145 days, as shown in Figure~\ref{tpt_optd}. In contrast, for normal Type IIP SNe, the interval between OPTd and $t_\mathrm{pt}$ is shorter, averaging $\sim$15.2$\pm$6.4 days. This estimated plateau duration for SN~2020aze aligns well with other known long plateau events, such as SN~2009ib ($t_\mathrm{pt} = 141 \pm 2$ days; \citealt{Takats2015_2009ib}), SN~2015ba ($t_\mathrm{pt} = 141 \pm 2$ days; \citealt{dastidar}), SN~2015an ($t_\mathrm{pt} = 130.4 \pm 0.3$ days; \citealt{Dastidr2019_2015an}), SN~2016B ($t_\mathrm{pt} = 133.5 \pm 0.6$ days; \citealt{Dastidr2019_2016B}), SN~2018hwm ($t_\mathrm{pt} = 130$ days; \citealt{Reguitti2021_2018hwm}), and SN~2021aai ($t_\mathrm{pt} = 140$ days; \citealt{Valerin2022_2021aai}). Since we lack observations during the radioactive tail phase, it is difficult to precisely constrain the amount of $^{56}$Ni synthesised in the explosion. Nevertheless, using the last photometric point during the transition phase in the $V$-band, we estimate an upper limit on the $^{56}$Ni mass of $\sim$0.04 M$_\odot$, based on the bolometric correction method of \cite{Hamuy2003}.

\begin{table}
	\centering
	\caption{Peak time, peak magnitudes and photospheric phase decline rate in the \textit{U}, \textit{B}, \textit{g$^\prime$}, \textit{V}, \textit{r$^\prime$}, \textit{i$^\prime$} bands.}
	\label{peak_mag}
    \begin{tabular}[width=\columnwidth]{c|c|c|c} 
	\hline
	Band & Epoch of maximum$^\dagger$ & Peak magnitude & Plateau decline$^\dagger$$^\dagger$\\
	     & (days)         & (mag) & rate (mag (100 day)$^{-1}$)\\
	\hline
        {\it U}  &  9.50$\pm$0.02 & $-$17.58$\pm$0.18 & 11.03$\pm$0.40\\
	{\it B}  &	9.82$\pm$0.36 & $-$16.70$\pm$0.17 & 4.91$\pm$0.47\\
	{\it g$^\prime$} &	12.11$\pm$0.22 & $-$16.83$\pm$0.17 & 3.04$\pm$0.13\\
	{\it V}  &	14.98$\pm$4.57 & $-$16.97$\pm$0.20 & 2.04$\pm$0.13\\
    {\it r$^\prime$} &	16.43$\pm$2.73 & $-$18.11$\pm$1.17 & 1.33$\pm$0.11\\
    {\it i$^\prime$} & 15.42$\pm$6.47 & $-$16.90$\pm$0.23 & 1.42$\pm$0.10\\
\hline
\end{tabular}
\newline
$^\dagger$Phase with respect to the explosion epoch (MJD = 58872.31).\\
$^\dagger$$^\dagger$ The decline rate is determined between 15 and 120 days post-explosion, or up to the last available photometric point if observations do not extend to 120 days.
\end{table}  

\begin{table*}
\begingroup
\footnotesize
\centering
\caption{Properties of the SNe in the comparison sample.}
\label{tab:comparison_objects}
\begin{threeparttable}
\adjustbox{center=\textwidth}{
\begin{tabular}{llcccccl}
\hline
Object    & Distance  & {\textit {E(B-V)}$_{tot}$} & t$_{pt}$  & s$_{50V}$ & M$_V^{peak}$ & M$_V^{50}$ & References\\
& (Mpc) &  (mag) & (days) & (mag/50 day) & (mag) & (mag) & \\
\hline
SN~2009au & 45.92 & 0.429$\pm$0.183 & - & 1.53$\pm$0.01 & $-$17.55$\pm$0.58 & $-$16.54$\pm$0.59 & 1\\ 
SN~2013fs & 50.95 & 0.05 & -
& 0.54$\pm$0.04$^*$ & $-$17.714$\pm$0.002$^*$ & $-$16.868$\pm$0.009$^*$ & 2\\
SN~2014G & 24.4$\pm$9.0 & 0.254$\pm$0.072 & $\sim$77 & 1.275 & $-$18.240$\pm$0.009$^*$ & $-$17.125$\pm$0.002$^*$ & 3\\
SN~2015bf & 33.89$\pm$0.05 & 0.148 & 56$_{-14}^{+10}$ & 1.22$\pm$0.09 & $-$18.11$\pm$0.08 & $-$17.085$\pm$0.002$^*$ & 4\\
SN~2017ahn & 40.76 & 0.264$\pm$0.054 & 57.56$\pm$0.88 & 2.04$\pm$0.11 & $-$18.44$\pm$0.29 & $-$16.591$\pm$0.020$^*$ & 6 \\
SN~2017gmr & 19.6$\pm$1.4 & 0.30 & 85 & 0.31$\pm$0.02$^*$ & $-$18.3 & $-$17.856$\pm$0.006$^*$ & 7, 8\\
SN~2018zd & 17.98$\pm$1.26 & 0.17$\pm$0.05 & 125 & 0.74$\pm$0.02$^*$ & $-$18.26$\pm$0.36 & $-$17.518$\pm$0.003$^*$ & 9\\
SN~2018lab & 35.5 & 0.22 & 113$\pm$3 & 0.13$\pm$0.05 & $-$15.1$\pm$0.1 & $-$14.957$\pm$0.002$^*$ & 10\\
SN~2021yja & 23.4$_{-4.4}^{+5.4}$ & 0.104 & - & 0.39$\pm$0.01$^*$ & $-$17.754$\pm$0.001$^*$ & $-$17.389$\pm$0.003$^*$ & 11\\
\textbf{SN~2020aze} & \textbf{36.7$\pm$2.6} & \textbf{0.20$\pm$0.02} & \textbf{$\sim$145} &  \textbf{1.02$\pm$0.07} & \textbf{$-$16.97$\pm$0.20} & \textbf{$-$15.89$\pm$0.17} & \textbf{This work}\\

\hline
\end{tabular}
}
\begin{tablenotes}

\setlength\labelsep{0pt}
\footnotesize{%
\item (1) \citet{Rodriguez2020}, (2) \citet{yaron2017}, (3) \citet{Bose2016}, (4) \citet{Lin2021}, (5) \citet{Reynolds2020}, 
(6) \citet{Tartaglia2017ahn}, (7) \citet{Andrews_2019}, (8) \citet{Utrobin2021}, (9) \citet{Zhang2020}, 
(10) \citet{Pearson}, (11) \citet{Hosseinzadeh}.\\
$^*$ Evaluated in this study.

}   
\end{tablenotes}
\end{threeparttable}
\endgroup
\end{table*}

\subsection{Comparison of SN~2020aze with other Type II SNe}

SN~2020aze is a luminous Type II SN with a bright peak, a fast-declining light curve, and an extended duration of photospheric phase. Its early spectra showed flash features containing narrow Balmer and \ion{He}{ii} lines, hinting at circumstellar interaction, before evolving into a standard Type II profile dominated by H$\alpha$. To place SN~2020aze in context, we assembled a comparison sample of Type II SNe that display early-time signatures of interaction with CSM, as evidenced by features in their spectra. The sample is selected based on literature reports of early-time flash-ionisation features and/or ledge features in the spectra. It includes six fast-declining events — SNe~2009au, 2013fs, 2014G, 2015bf, 2017ahn, and 2018zd — as well as two slow-declining SNe~2017gmr and 2021yja, and a low-luminosity SN~2018lab. SN~2009au also exhibits a long photospheric phase akin to that of SN~2020aze.  The parameters of these comparison SNe are listed in Table~\ref{tab:comparison_objects}.

\begin{figure}
\includegraphics[width=\columnwidth]{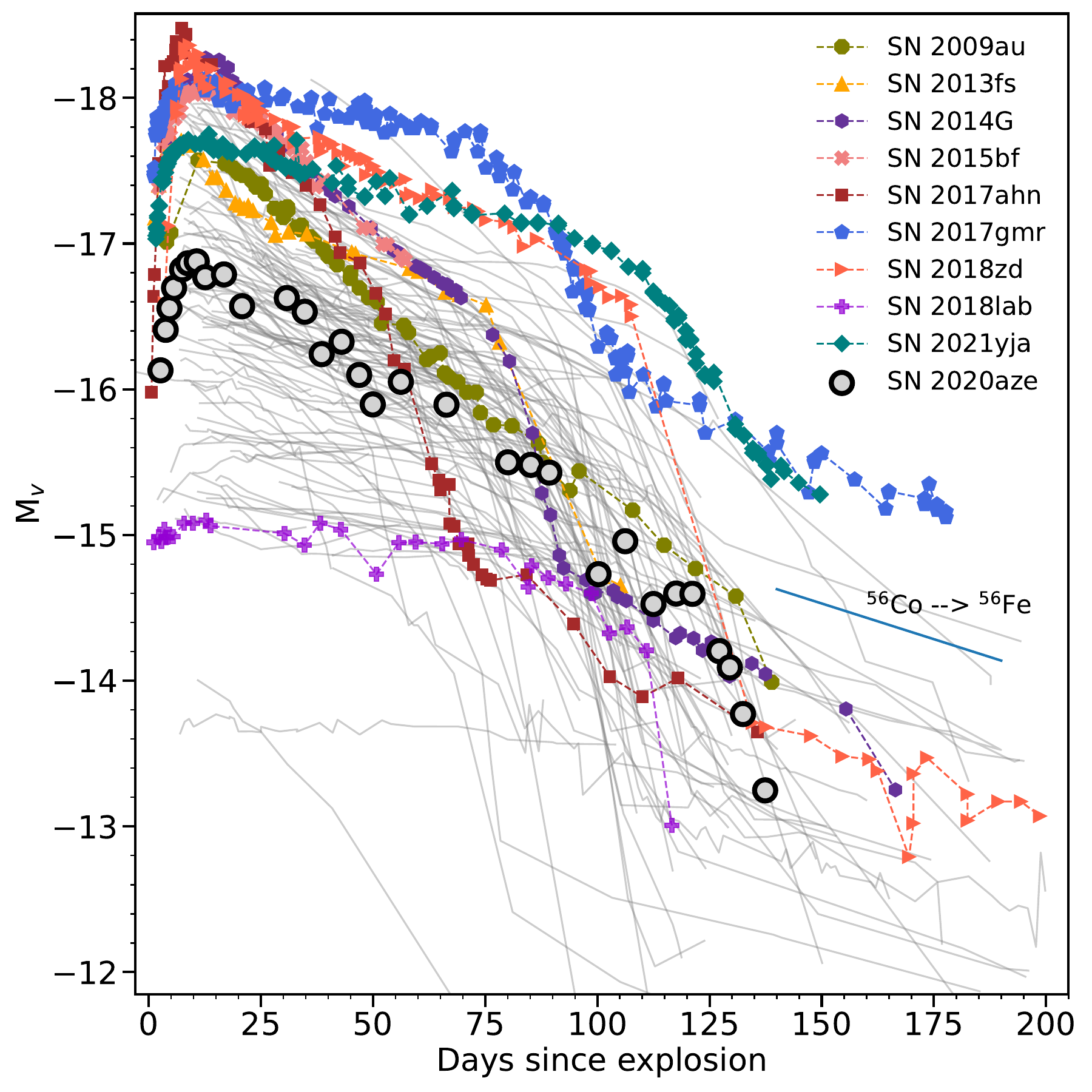}
    \caption{Comparison of the $V$-band absolute magnitude light curve of SN~2020aze with other Type II SNe. The grey light curves represent the 116 Type II SNe with varying photospheric phase durations taken from \citet{anderson2014}.}
    \label{absmag}
\end{figure}

Figure~\ref{absmag} compares the absolute V band light curve of SN~2020aze with other SNe in the comparison sample. The t$_{\rm pt}$ and peak magnitude of SN~2020aze are 145 days and $-$16.97$\pm$0.20 mag, respectively. SNe with early flash features have a brightness similar to or greater than SN~2020aze except SN~2018lab, which is a low-luminous Type II SN. The decline rate (s$_2$) in the photospheric phase of SN~2020aze is less steep (2.04$\pm$0.13 mag (100 day)$^{-1}$) compared to SN~2014G (2.55 mag (100 day)$^{-1}$). SN~2009au, a luminous with low-expansion velocity (LLEV; \citealt{Rodriguez2020}) SN, shows a similar peak magnitude ($-$17.55$\pm$0.58 mag) as SN~2020aze, though the decline rate is steeper (3.06 mag (100 days)$^{-1}$) than SN~2020aze.

\begin{figure}
    \includegraphics[width=\columnwidth]{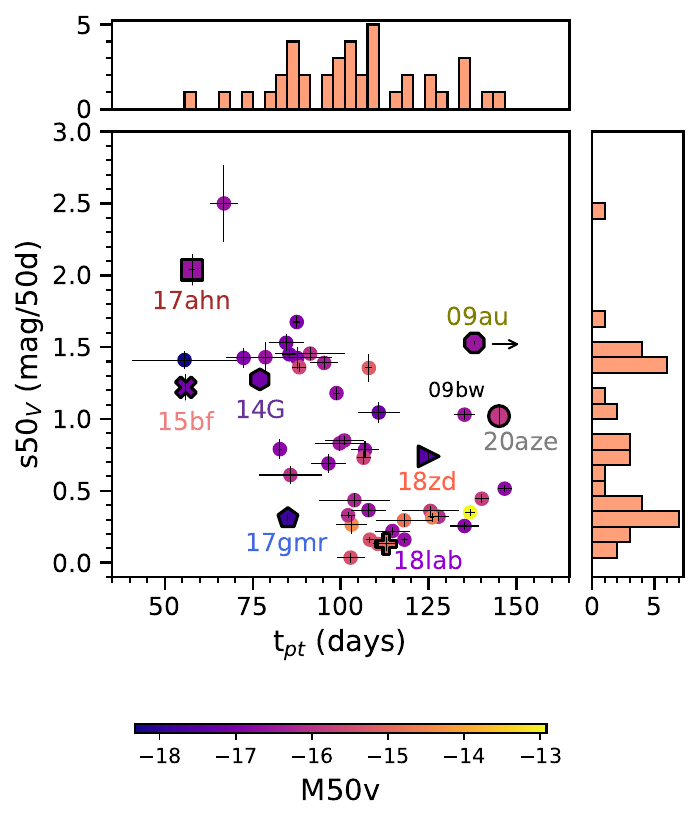}
    \caption{Correlation between the slope of $V$-band light curve (s50$_V$) and the plateau duration (t$_{\rm pt}$) is shown. The data points are colour-coded with the magnitude at 50 day (M$^{50}_V$). The location of SN~2020aze is highlighted.}
    \label{sv50_tpt_mv}
\end{figure}

Based on these light curve parameters, we tried to co-locate SN~2020aze in the parameter space of Type II SNe from \citet{anderson2014} and \citet{valenti16}. In Figure~\ref{sv50_tpt_mv}, the post-peak decline rate of the $V$-band light curve over 50 days (s50$_V$) is plotted as a function of the plateau duration (t$_{\rm pt}$), colour-coded with the $V$-band absolute magnitude at day 50. Typically, faster-declining events exhibit a brighter peak and a shorter photospheric phase \cite{anderson2014, valenti16}. However, SN~2020aze deviates from this trend, showcasing a luminous peak, with a fast-decline but an extended photospheric phase duration similar to SN~2009au. Although SN~2009bw lies closest to SN~2020aze in s50$_V$, it exhibits a comparatively shorter photospheric phase. 

\subsection{Colour evolution}
\begin{figure}   
\includegraphics[width=\columnwidth]{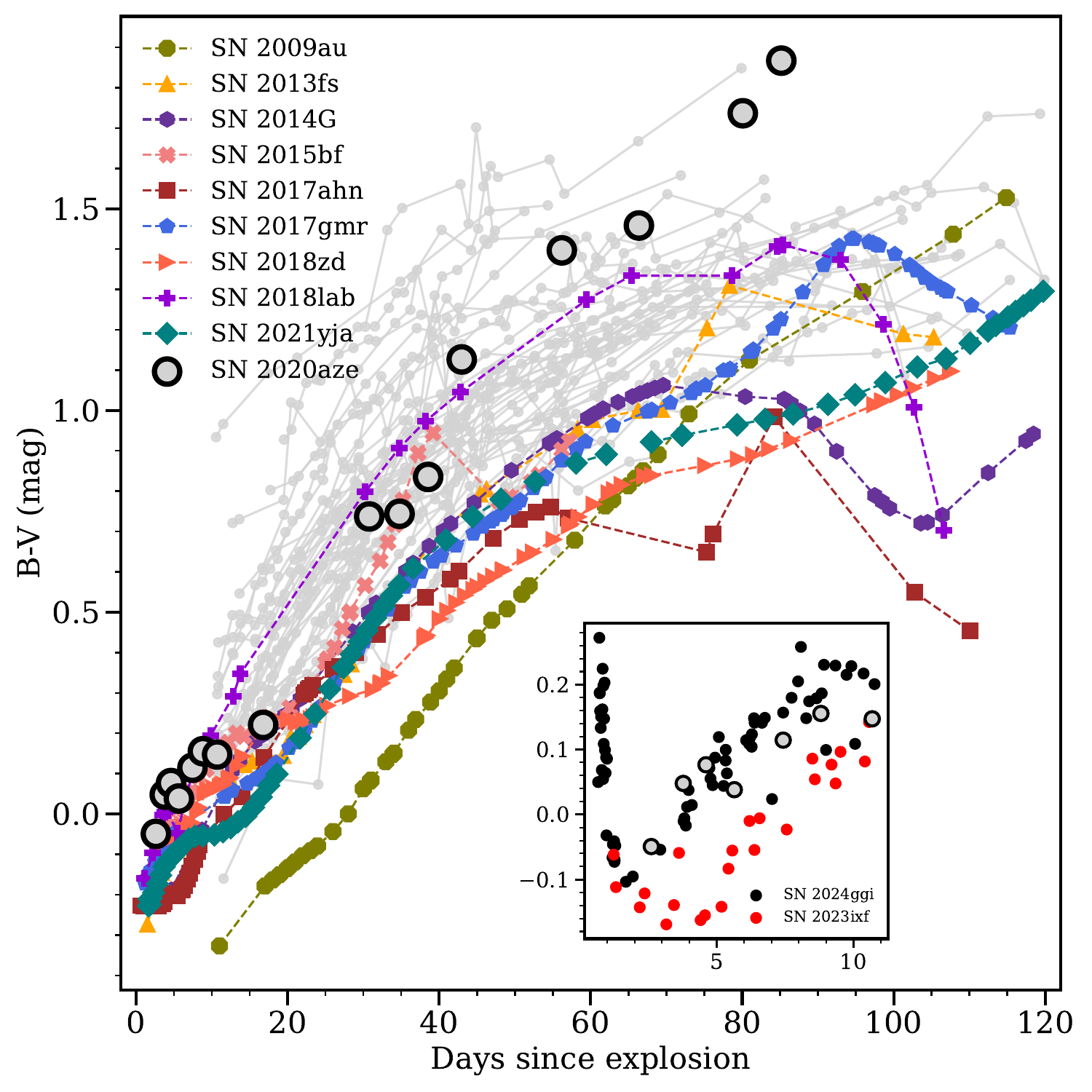}
    \caption{Comparison of ($B-V$) colour evolution of SN~2020aze with other Type II SNe. Grey lines in the background represent the sample from \citet{dejaeger2018}.}
    \label{B-Vf}
\end{figure}

The reddening-corrected ($B-V$) colour evolution of SN~2020aze, along with its comparison to other SNe, is presented in Figure~\ref{B-Vf}. Before calculating the colour, the $B$ and $V$ band light curves were first smoothed using the \texttt{Locally Weighted Regression (LOESS; \citealt{Cleveland_loess})} technique, with a smoothing parameter of 0.2, which defines the fraction of data points used for local regression. The resulting colour evolution shows consistency across all the SNe studied. As the recombination phase begins, the colour becomes progressively redder due to the expansion and cooling of the SN ejecta. Notably, during the recombination phase, the ($B-V$) colour of SNe~2018zd and 2021yja exhibits a flat region, whereas the ($B-V$) colour of SN~2020aze continues to become redder. SN~2020aze is redder during the recombination phase as compared to other SNe in the comparison sample, except for SN~2018lab, which is a low-luminosity object. As the photospheric phase ends, the colour of SNe~2014G, 2017ahn, and 2018lab becomes bluer. Due to limited observations, we were unable to capture the complete fall from the photospheric phase to the radioactive tail in SN~2020aze. In the inset plot of Figure~\ref{B-Vf}, the early colour evolution of SN~2020aze is compared with those of SNe~2023ixf and 2024ggi. Both of these SNe show a rapid evolution toward bluer colours in the first few days, interpreted as arising from interaction with a dense CSM, followed by a turnover in the colour evolution. The blue maximum occurs at 4 days for SN~2023ixf \citep{azalee_2023ixf} and 2 days for SN~2024ggi \citep{shrestha2024}. SN 2020aze colours are similar to SN~2024ggi, though the absence of early observations prevents us from confirming any initial blueward evolution.

\section{Light Curve Modelling}
\label{LC modelling}

\subsection{Early Light Curve Modelling}
\label{shockcooling}

\begin{table*}
\begingroup
\footnotesize
\centering
\caption{Parameters of the best fit shock cooling model.}
\label{tab:shock_cooling}
\begin{threeparttable}
\resizebox{\textwidth}{!}{
\begin{tabular}{lcccccc}
\hline
Parameter    & Variable  & \multicolumn{3}{c}{Prior} & Best-fit Values & Units\\
\cline{3-5}
& & Shape & Min. & Max. & & \\

\hline
Shock Velocity & v$_{s*}$ & Uniform & 0.0 & 5.0 & 0.7$_{-0.1}^{+0.2}$ & 10$^{8.5}$ cm/s\\
Envelope Mass & M$_{env}$ & Uniform & 0.0 & 10.0 & 7.0$_{-3.0}^{+3.0}$ & M$_\odot$\\
Ejecta Mass$\times$numerical factor & f$_{\rho}$ M & Uniform & 0.01 & 100.0 & 40.0$_{-30.0}^{+40.0}$ & M$_\odot$\\
Progenitor Radius & R & Uniform & 0.7 & 10 & 3.2$_{-0.9}^{+1.0}$ & 10$^{13}$ cm\\
Explosion Time & t$_{0}$ & Uniform & 58872.0 & 58874.0 & 58873.4$_{-0.4}^{+0.3}$ & MJD\\ 
Intrinsic Scatter & $\sigma$ & Uniform & 0.0 & 100.0 & 9.0$\pm$1.0 & --\\

\hline
\end{tabular}
}
\end{threeparttable}
\endgroup
\end{table*}

SN explosions occur with a short flash of high-energy radiation followed by optical/NIR emission from the expanding layers of SN ejecta. The early-phase light curve analysis of CCSNe aids in estimating the parameters of the SN explosion and its progenitor star \citep{sapirwax2017, Morag2023}. We modelled the early-phase light curve of SN~2020aze using the Markov Chain Monte Carlo (MCMC) code developed by \cite{griffin2023}, based on the shock-cooling model of \citet{Morag2023} (\texttt{MSW23}). The \texttt{MSW23} model is optimised for progenitors characterised by low CSM column density and optical depth, where radiation is driven by the expanding stellar envelope, making it an ideal choice for SN~2020aze. The model estimates the following parameters: shock velocity  (v$_s$*), envelope mass (M$_{env}$), ejecta mass $\times$ numerical factor (f$_{\rho}$M), progenitor radius (R), and explosion epoch (t$_0$). In addition, we also include an intrinsic scatter term, $\sigma$, which effectively increases the observed error bars by a factor of $\sqrt{1 + \sigma^2}$. This model assumes a polytropic index of $n=1.5$ for the RSG progenitor density profile.

We fit the early multi-band $\it UVM2$, $\it UVW2$, $\it UVW1$, $UBVg^\prime r^\prime i^\prime$ light curves of SN~2020aze with the shock-cooling model up to MJD 58879, which corresponds to $\sim$7 day since explosion, considering uniform priors for all variables. The selection of epochs is based on the shock-cooling criteria described in Equation A3 of the Appendix in \cite{Morag2023}. Figure \ref{posterior} shows the MCMC fit of the light curve to the observed data points. For each parameter set, the model generates the total luminosity and blackbody temperature of the SN over time. These are then transformed into fluxes for each photometric point, over which light curve fitting is performed. The model constrains the explosion time to be MJD = 58873.4 day, which is within errors of the estimated explosion epoch in Section~\ref{sn2020aze_properties}. The best fit parameters are listed in Table~\ref{tab:shock_cooling}. While the fits to the optical light curves are generally well-matched, they fail to capture the UVOT bands. The fit yields a progenitor radius of approximately 3.2$\times$10$^{13}$ cm = $\sim$500 R$_\odot$, which is within the typical range for RSG stars (100–1500 R$_\odot$; \citealt{Massey1998, Massey2003, Levesque2017, Neugent2020}). The parameter f$_{\rho}$M is estimated to be around 40.0 M$_\odot$; however, the early light curve is not strongly influenced by f$_{\rho}$M, making this estimate uncertain \citep{Morag2023}. Additionally, the best-fit shock velocity is lower than the measured photospheric velocities, which is physically inconsistent. As the shock-cooling model fails to match the observations across all bands, the derived parameters cannot be considered reliable.

We note that our observations begin at $\sim$3.6 days post-explosion, a phase where standard shock-cooling models are typically expected to be valid. However, the \citet{Morag2023} formalism fails to capture the UV evolution of SN~2020aze at these epochs, suggesting a sustained UV excess that cannot be explained by envelope cooling alone. While \citet{Irani_classi} identified model failures in the first 24–48 hours due to confined CSM, our results may imply the presence of a more extended circumstellar environment. This interaction likely provides an additional luminosity source that continues to power the light curve and influence the UV evolution well into the cooling phase, leading to the discrepant model parameters discussed above.

\begin{figure} \includegraphics[width=\columnwidth]{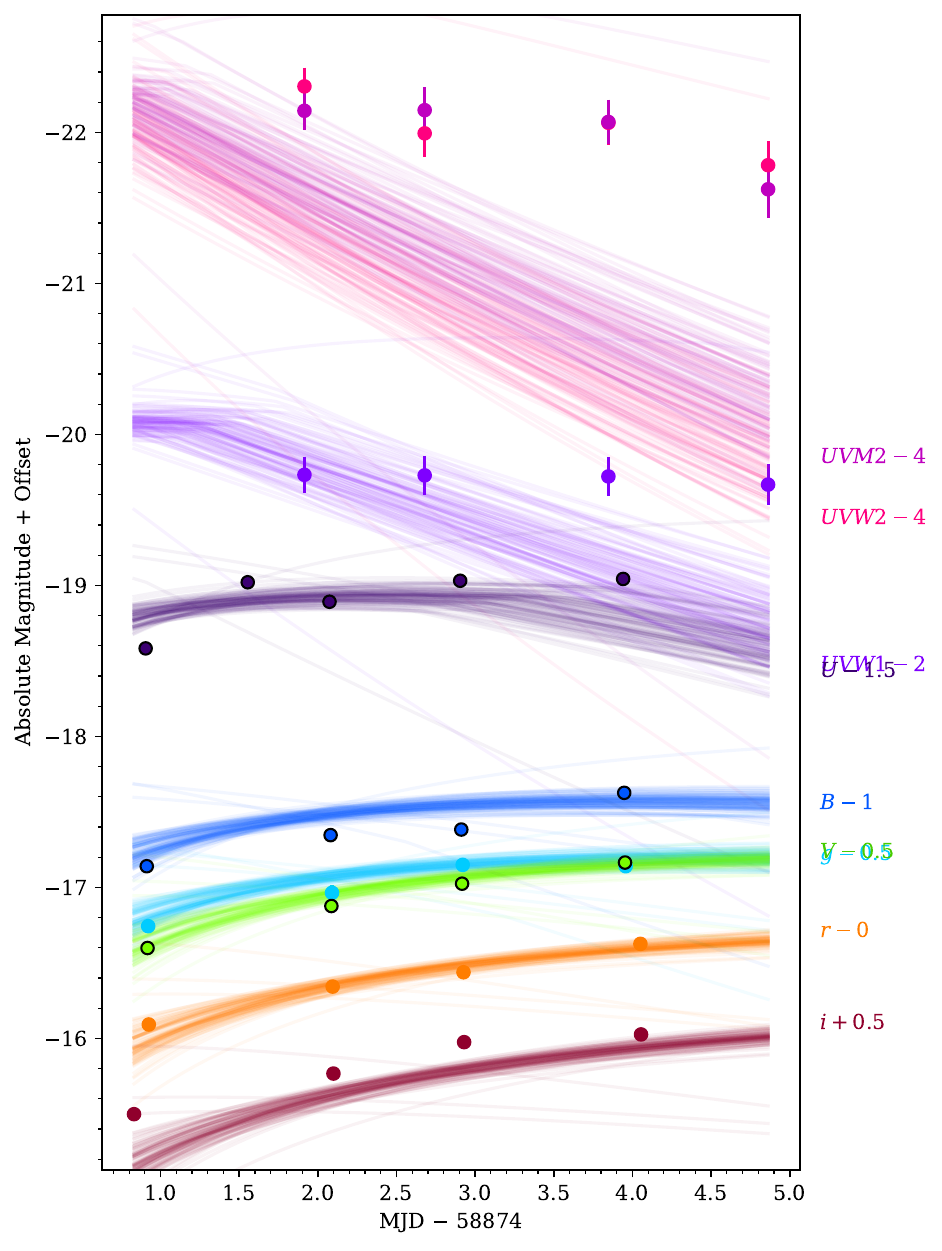}
    \caption{Light-curve modelling of SN~2020aze using the MCMC code by \citet{griffin2023}, based on the shock-cooling framework of \citet{Morag2023}. The model accurately reproduces the observed light curves across all photometric bands.}
    \label{posterior}
\end{figure}

\subsection{Bolometric light curve modelling}
\label{nagi_vinko}

\begin{figure}
\includegraphics[width=\columnwidth]{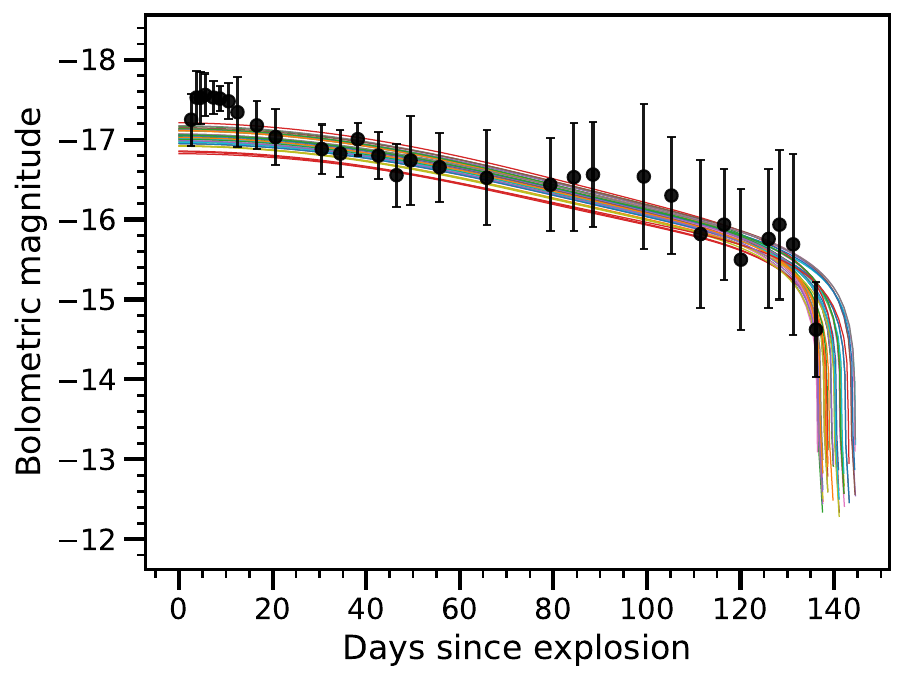}
    \caption{Evolution of the bolometric light curve of SN~2020aze is plotted along with 45 best fitted MCMC light curves using the semi-analytical modelling prescription of {\protect\cite{Nagyvinko2016}} and {\protect\cite{Jager}}.}
    \label{nagi_vinko_plot}
\end{figure}

We performed semi-analytical modelling of the bolometric light curve to obtain estimates of the SN explosion parameters such as the initial ejecta radius (R$_0$), ejecta mass (M$_{\rm ej}$), initial kinetic energy (E$_{\rm kin}$), and thermal energy (E$_{\rm th}$). The bolometric light curve displayed in Figure~\ref{nagi_vinko_plot} was generated using the {\it SDAUBVg}$^\prime${\it r}$^\prime${\it i}$^\prime$ bands up to 11 days, then {\it UBVg}$^\prime${\it r}$^\prime${\it i}$^\prime$ up to 46 days, and thereafter with the {\it BVg}$^\prime${\it r}$^\prime${\it i}$^\prime$ bands, utilizing the \texttt{SUPERBOL} code \citep{nicholl}. The S, D, and A filters correspond to the UVOT UVW2, UVM2, and UVW1 bands, respectively. \texttt{SUPERBOL} interpolates magnitudes in all filters to the epochs of a reference filter and converts them to flux values, which are then used to compute the photometric spectral energy distribution (SED) at each epoch. The SED is fitted with a blackbody and extrapolated to the UV and NIR regimes, enabling the determination of the bolometric luminosity at all epochs.

\begin{table}
	\centering
	\caption{Best-fitted explosion parameters along with 1$\sigma$ confidence interval from the bolometric lightcurve of SN~2020aze \citep{Nagyvinko2016}.}
	\begin{tabular}{lll} 
        \hline
        Parameter & Variable & Best-fit value \\
        \hline
		Initital radius of ejecta &  R$_0$ (10$^{13}$ cm) & 7.60$\pm$1.05\\
            Ejecta Mass & M$_{\rm ej}$ (M$_\odot$) & 12.24$\pm$0.23\\
            Initial kinetic energy & E$_{\rm kin}$ ($10^{51}$ erg) & 1.14$\pm$0.07\\
            Initial thermal energy & E$_{\rm th}$ ($10^{51}$ erg) & 0.41$\pm$0.06\\
            \hline
	\end{tabular}
	\label{tab:nagyvinko}
\end{table}

The bolometric light curve is modelled using the semi-analytical model of \cite{Nagyvinko2016}. The model incorporates a dense core and an extended outer shell, usually called the two-component model. The two-component approximation allows us to fit the early decline in the light curve with the shell component and the photospheric plus the radioactive tail phase with the core component. Later, the MCMC technique was integrated into this by \cite{Jager}, which extracts the best fit through $\chi^2$ minimisation for the observed bolometric light curve and provides the best fit physical parameters. The implementation by \cite{Jager} only considers the core component and cannot fit the early light curve. Therefore, the fitting is performed starting from 20 days after the explosion. The 45 best-fit model light curves to the bolometric luminosities of SN~2020aze are shown in Figure~\ref{nagi_vinko_plot}. These light curves were selected because they capture the transition from the photospheric phase to the radioactive tail phase within 140 days post-explosion (marking the onset of the tail phase). Table~\ref{tab:nagyvinko} lists the best-fit estimates of the physical parameters alongside their prior ranges. These parameters represent the mean values derived from these 45 light curves. The estimated ejecta mass of 12.2 M$_{\odot}$ is different from the value obtained from the early light curve modelling discussed in Section~\ref{shockcooling}. This difference arises because the early light curve is not fully sensitive to the total ejecta mass when luminosity from the interaction dominates. Combining with the mass of the nascent neutron star (1.5-2 M$_{\odot}$) results in a progenitor mass of around 14 M$_{\odot}$ at explosion. The progenitor radius is estimated to be approximately 7.6$\times$10$^{13}$ cm ($\sim$1100 R$_{\odot}$), which is within the limit of RSG stars.

\section{Spectral Analysis}
\label{spectra}

\begin{figure}
\includegraphics[width=\columnwidth]{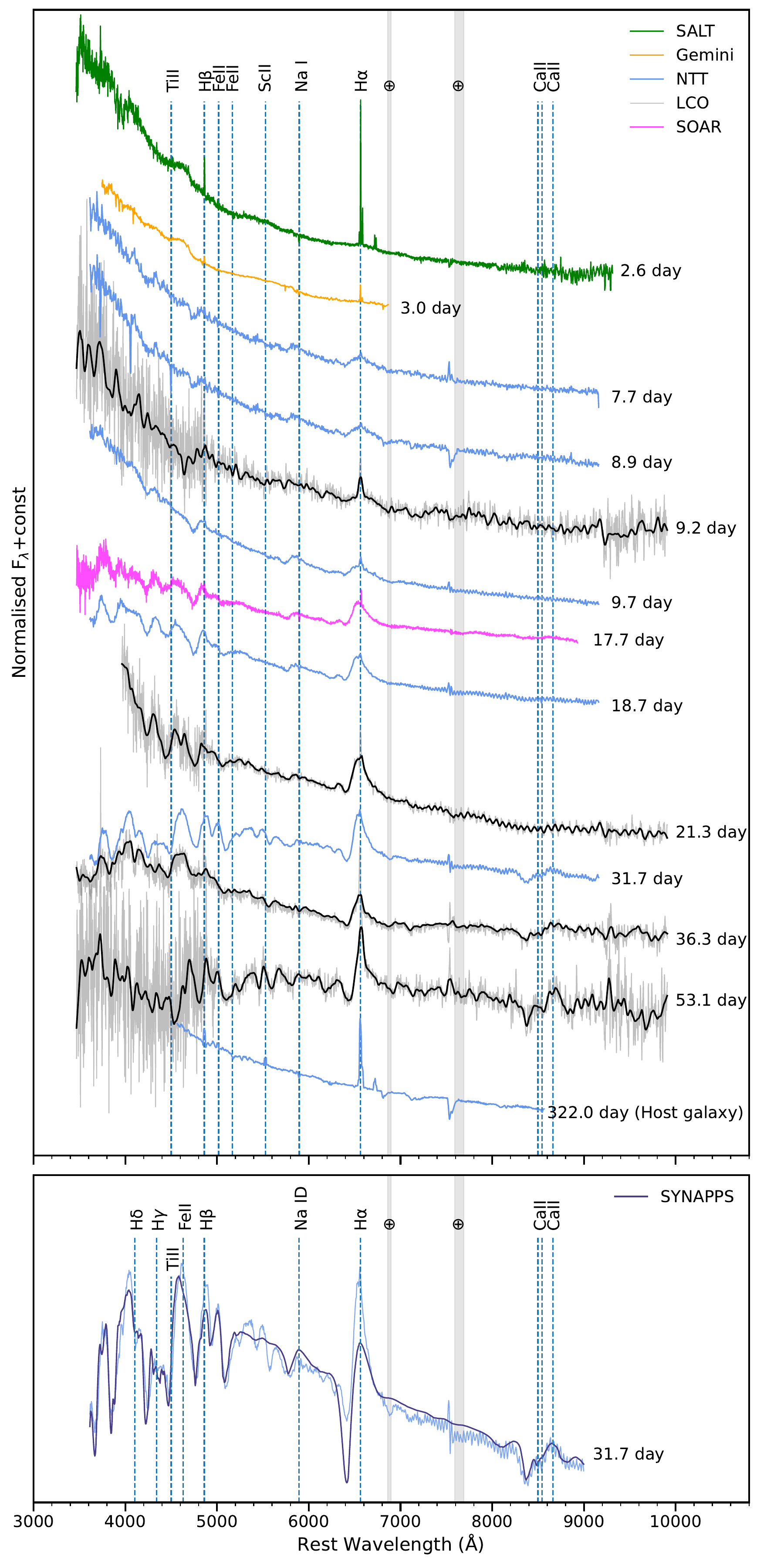}
\caption{Spectral evolution of SN~2020aze from 2.6 to 322.0 days since the explosion, with the identified lines shown in the upper panel. The spectra are corrected for redshift and extinction. The actual spectra (grey) are smoothed (black) for line identification. The lower panel displays the best-fit models from \texttt{SYNAPPS} for the spectra at 31.7 days. The spectral features identified through the modelling are also marked, while the grey band in both plots indicates regions affected by telluric absorption.}
\label{Spectral_evolution}
\end{figure}

The dereddened spectral sequence of SN~2020aze is presented in the upper panel of Figure~\ref{Spectral_evolution} from 2.6 to 322.0 day since the explosion. The first two spectra display a blue continuum with a few narrow lines arising from the host galaxy. In addition, a bump spanning 4400 to 4800 \AA{} is observed, which disappears thereafter. The presence of the \ion{Na}{i} D absorption line at the redshift of the host gives an idea of the host galaxy extinction (as discussed in Section~\ref{sn2020aze_properties}). The final spectrum, obtained at 322.0 days, is dominated by the host galaxy emission, with a prominent H$\alpha$ line.

We compare the observed spectrum of SN~2020aze in the photospheric phase (31.7 day) with the synthetic spectrum generated with the parameterised spectrum synthesis code \texttt{SYNAPPS} \citep{Thomas_synapps_2011}. This code assumes spherical symmetry, homologous ejecta expansion, local thermodynamic equilibrium and Sobolev's approximation for line formation \citep{Sobolev}. The synthetic spectrum is generated using the \ion{H}{i}, \ion{Na}{i}, \ion{Fe}{i}, \ion{Fe}{ii}, \ion{Ca}{ii}, \ion{Sc}{ii} and \ion{Ti}{ii} ionisation states and is shown in the lower panel of Figure~\ref{Spectral_evolution}. The photospheric velocity in the best-fit model is 6200 km s$^{-1}$ and the outer ejecta velocity used in the modelling is 30,000 km s$^{-1}$. The photospheric temperature in the best-fit model is 6900 K. 

\begin{figure}    
\includegraphics[width=\columnwidth]{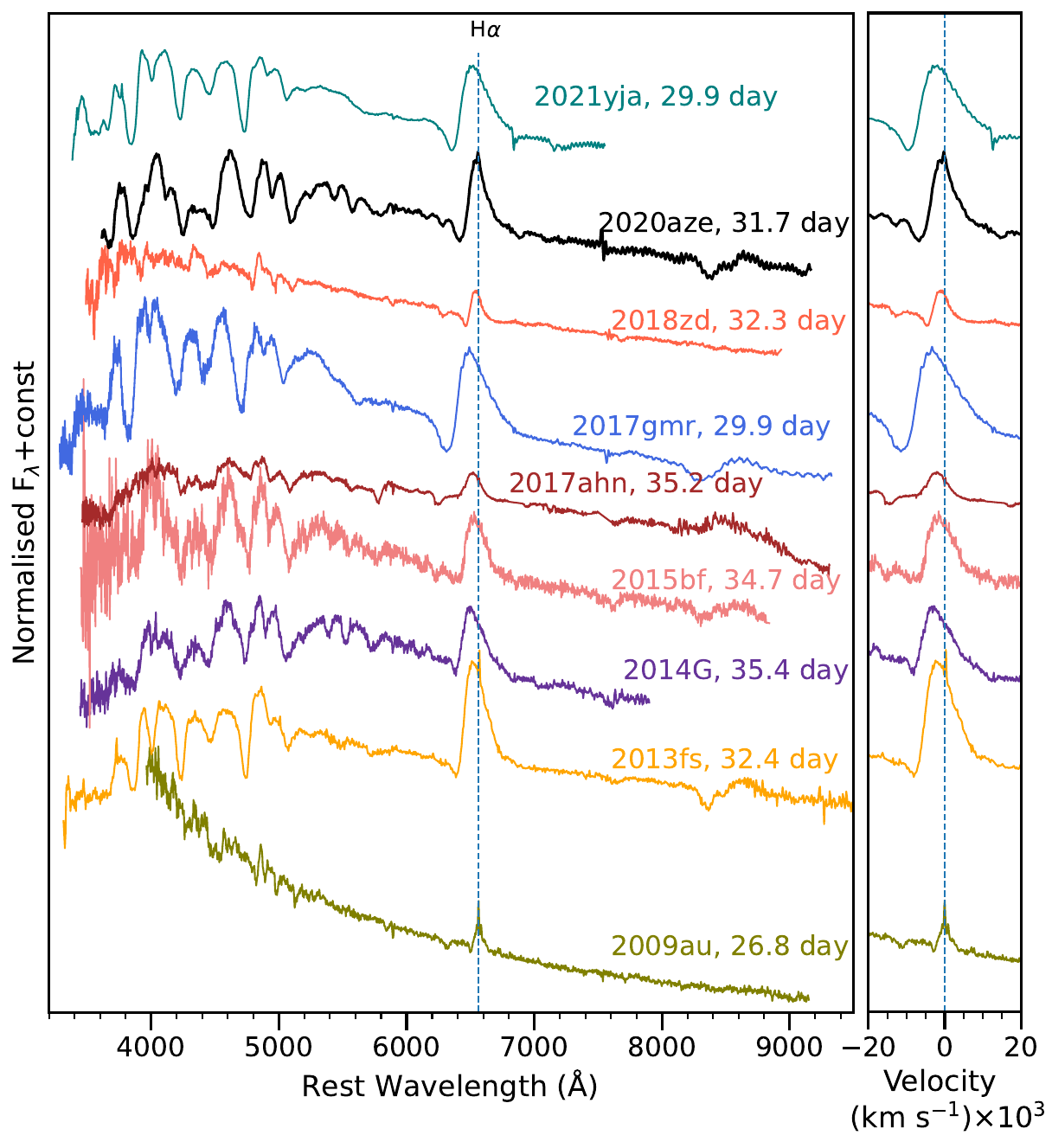}
    \caption{Comparison of the SN~2020aze spectrum with other Type II SNe at 31.7 day. The H$\alpha$ line profile in velocity space along the x-axis is also indicated.}
    \label{IIPL}
\end{figure}

In the photospheric phase spectra, various metal lines like the NIR \ion{Ca}{ii} $\lambda\lambda\lambda$8498, 8542, 8662, \ion{Sc}{ii} $\lambda$5527, and \ion{Fe}{ii} $\lambda\lambda$5018, 5169 features start appearing but the strength of these lines are not prominent. The absorption feature of H$\alpha$ P-Cygni profile in the photospheric phase is not very pronounced in the case of SN~2020aze. Figure~\ref{IIPL} shows the spectral comparison of SN~2020aze with other Type II SNe. The P-Cygni profile in fast-declining Type II SNe, such as SNe~2015bf and 2017ahn, does not show the absorption dip. SNe~2014G and 2018zd have a weak P-Cygni profile. The slow declining SNe~2013fs, 2017gmr, and 2021yja have a prominent absorption dip of the H$\alpha$ line in their spectra. From the Type II sample study of \cite{Gutirrez2014}, it was found that a lower a/e value (absorption to emission ratio) of the H$\alpha$ profile is correlated with a faster decline rate and shorter photospheric phase. For SN~2020aze, the a/e ratio for spectra around 32 days is approximately 0.26. Such a low value is consistent with a faster decline rate but does not correspond to a shorter photospheric phase duration. The extended photospheric phase could be due to a progenitor with an extended hydrogen envelope, while the presence of a dense CSM may contribute to the steeper light curve decline \citep{Morozova2017}.

\subsection{Early interaction signatures}

\begin{figure}
    \includegraphics[width=\columnwidth]{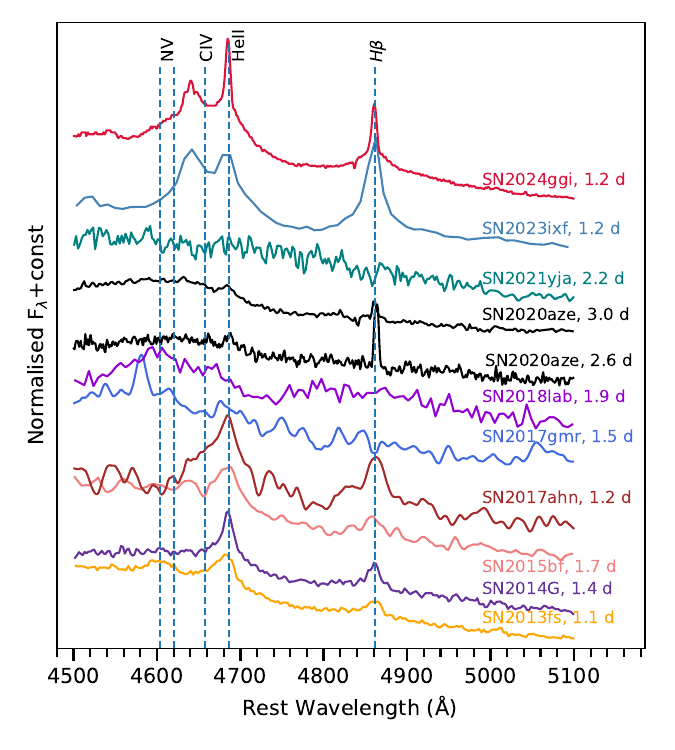}
    \caption{Comparison of early flash features visible in SN~2020aze and other Type II SNe. }
    \label{flash}
\end{figure}

In the 2.6 and 3.0 day spectrum of SN~2020aze, a distinct bump spanning the wavelength range from 4400 to 4800 \AA{} is apparent but disappears in later phases. Narrow high-ionisation emission features like \ion{He}{ii}, \ion{Ca}{iii}, \ion{Ca}{iv}, \ion{N}{iv}, and \ion{O}{iv} are commonly observed in Type II SNe discovered shortly after explosion, such as SNe~2023ixf and 2024ggi. These features persist until the ejecta completely engulfs the CSM, offering valuable insights into the density, composition, and velocity of the CSM surrounding the progenitor star \citep{Boian2019}. However, in some Type II SNe, such as SNe~2017gmr, 2018lab, and 2021yja, a ledge-like feature (bump) is observed instead of narrow emission lines, resembling the ledge feature in the early spectrum of SN~2020aze. A comparison of early interaction signatures, in the form of narrow and broad emission features in Type II SNe, is shown in Figure~\ref{flash} alongside SN~2020aze. 

\begin{figure}
    \includegraphics[width=\columnwidth]{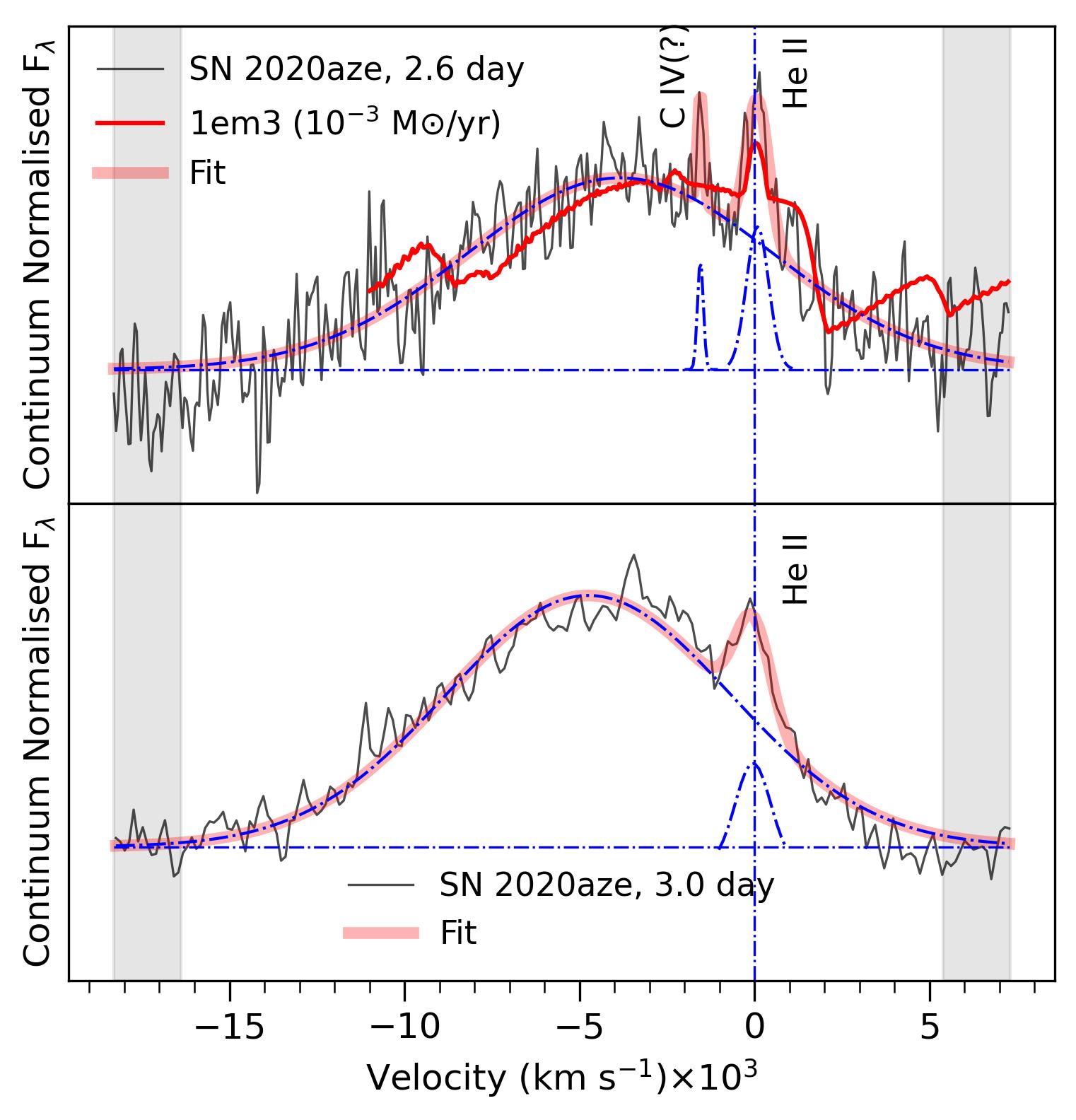}
    \caption{Fit to the spectral region around \ion{He}{ii} $\lambda$4686 showing the presence of a ledge feature. In the 2.6 d spectrum, the composite profile is modelled with two narrow Gaussian components and one broad Gaussian component (shown in the top panel), whereas in the 3.0 d spectrum it is modelled with one narrow and one broad component (shown in the bottom panel). The zero velocity corresponds to the rest velocity of \ion{He}{ii} $\lambda$4686. The \texttt{1em3} model spectrum from \citep{Dessart2023} corresponding to 2 day after explosion, is over-plotted in red in the top panel.}
    \label{ledge}
\end{figure}

The appearance of the ledge-like bump may originate from the blending of several ionised CSM features \citep{Soumagnac2020}, radiative acceleration of the CSM \citep{Tsuna2023}, blue-shifted \ion{He}{ii} $\lambda4686$ emission produced by SN ejecta interacting with a surrounding CSM shell \citep{Anderson2012}, or dense clumps formed at the ejecta boundary are heated by the forward shock \citep{Chugai2023_HeII}. To investigate this feature in SN~2020aze, we fit the early spectra with multiple Gaussian components (Figure~\ref{ledge}). The 2.6-day spectrum can be decomposed into a broad Gaussian centred at 4626 \AA{} together with two narrow Gaussians at 4662 \AA{} and 4687 \AA{}; the narrow lines coincide with the rest wavelengths of \ion{C}{iv} $\lambda4658$ and \ion{He}{ii} $\lambda4686$, with FWHM velocities of $\sim$200 km s$^{-1}$ and $\sim$750 km s$^{-1}$, respectively, while the broad feature reaches $\sim$10,000 km s$^{-1}$ if attributed to blue-shifted \ion{He}{ii}. The 3.0-day spectrum shows the same broad component at 4626 \AA{} and a single narrow feature at 4687 \AA{}, with corresponding FWHM velocities of $\sim$9,600 km s$^{-1}$ (ledge) and $\sim$1,200 km s$^{-1}$ (\ion{He}{ii}).  Although a potential \ion{C}{iv}$\lambda$4658 contribution is noted at 2.6 days, the instrumental resolution (R$\approx$1100) corresponds to a velocity resolution of $\sim$270~km~s$^{-1}$, rendering this identification uncertain. Furthermore, the absence of the corresponding \ion{C}{iv} $\lambda$5801, 5812 doublet in the 2.6 days suggests that this feature is inconsistent with a \ion{C}{iv} identification. Since narrow emission components are observed in addition to the broad feature in SN~2020aze, it is unlikely that the bump results from broadening of the narrow features. While radiative acceleration of the CSM can produce blue-shifted broadened features, as proposed in SN~2024ggi \citep{Pessi}, these typically exhibit velocities around $\sim$1,000 km s$^{-1}$ and cannot explain the extreme velocities observed in the case of SN~2020aze.

The most plausible explanation for the ledge-like feature in SN~2020aze is blue-shifted \ion{He}{ii} $\lambda$4686, produced in the outer layers of SN ejecta with an expansion velocity of $\sim$10,000 km s$^{-1}$. The intermediate \ion{He}{ii} feature with velocities close to 1,000 km s$^{-1}$ likely corresponds to intermediate-width wings, believed to result from thermal electron scattering of the narrow-line emissions \citep{Chugai2001, Dessart2009, Huang2018}.     

\cite{Dessart2023} investigated the spectral characteristics of RSG explosions using the NLTE radiative-transfer model, \texttt{CMFGEN}, focusing on scenarios where the progenitor is surrounded by CSM. They considered explosions of RSGs having progenitor radii (R$_\star$; 501 R$_\odot$), accounting for mass-loss rates between 10$^{-5}$ -- 10$^{0}$ M$_\odot$ yr$^{-1}$, while maintaining a constant ejecta mass of 12.52 M$_\odot$ and CSM velocity of 50 km s$^{-1}$. We compared the early spectrum of SN~2020aze at 2.6 day with the models and the optimum match was found with 2.0 day epoch of \texttt{1em3} model, which is shown in the top panel of Figure~\ref{ledge}. This model approximates the progenitor as having a wind mass-loss rate of $\sim$10$^{-3}$ M$_\odot$ yr$^{-1}$. The resulting low-density CSM is optically thin, which explains the absence of pronounced narrow flash features in the modelled spectrum or it might be the doppler-broadened \ion{He}{ii} line. 

\begin{figure}
    \includegraphics[width=\columnwidth]{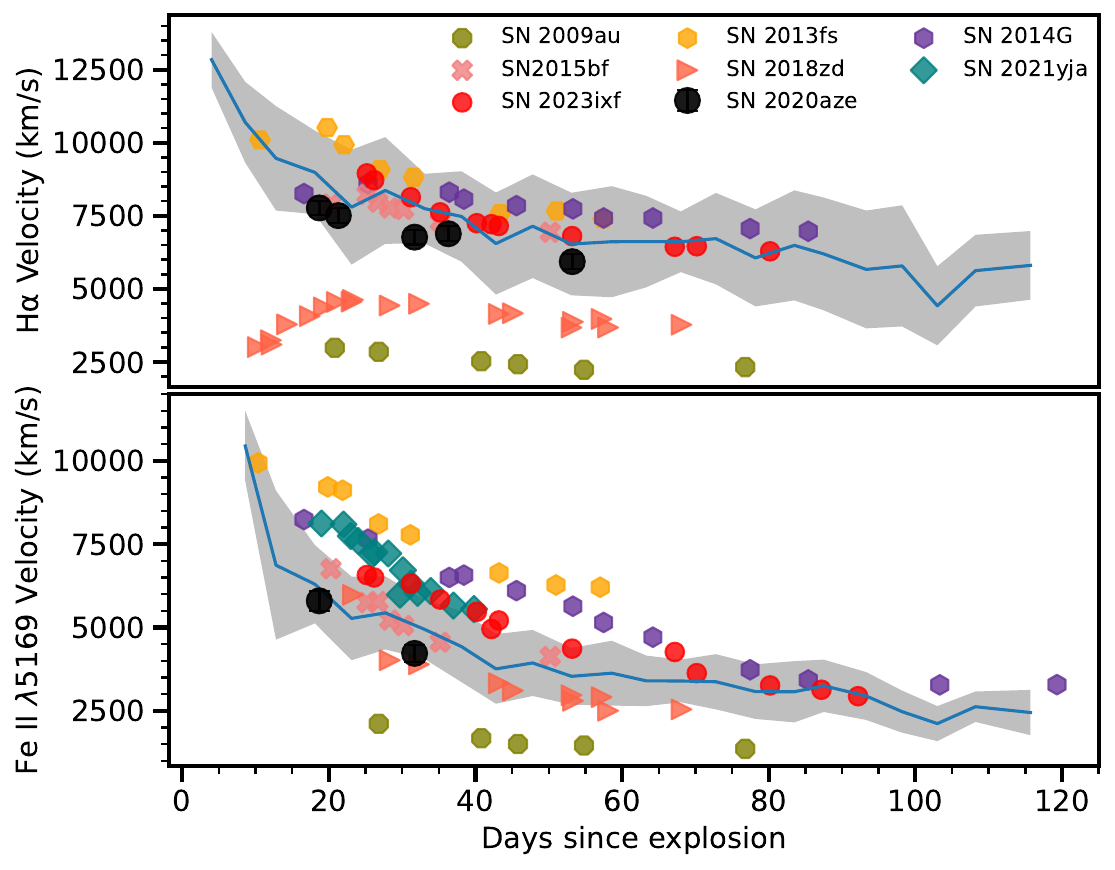}
    \caption{Velocity evolution of SN~2020aze and comparison with other SNe. The mean velocities of the Type II SNe sample from \citet{gutirrez2017a} are indicated by a blue solid line, while the grey band represents the corresponding standard deviations.}
    \label{velocity}
\end{figure}

\begin{figure}
    \includegraphics[width=\columnwidth]{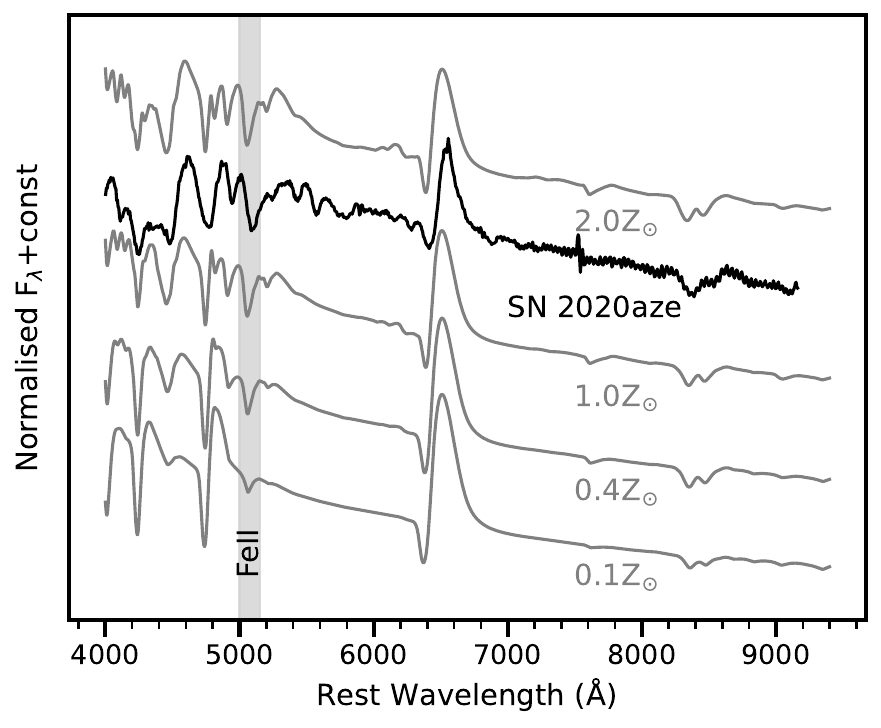}
    \caption{Comparison of the 30 day deredshifted and reddening corrected spectrum of SN~2020aze with the model spectra of \citet{Dessart2014} at different metallicities.}
    \label{metallicity}
\end{figure}

\subsection{Expansion velocity and progenitor metallicity}
The velocity evolution of the H$\alpha$ and \ion{Fe}{ii} $\lambda$5169 profiles of SN~2020aze, along with other SNe, is shown in Figure~\ref{velocity}. The velocity is determined through the absorption dip of the P-Cygni profile across two prominent species (H$\alpha$ and \ion{Fe}{ii} $\lambda$5169). Given the restricted spectroscopic data available for SN~2020aze, the velocity is calculated in the last five spectra. Due to low-quality spectra, the \ion{Fe}{ii} line velocity can only be estimated at two epochs. The H$\alpha$ and \ion{Fe}{ii} velocity of SN~2020aze falls below the mean velocity level derived from a sample of 122 Type II SNe presented in \cite{gutirrez2017a}, as shown in the upper panel of Figure~\ref{velocity}. Still, it remains within one standard deviation, as illustrated by the shaded grey region. The \ion{Fe}{ii} $\lambda$5169 velocities for all SNe with flash features are at or above one standard deviation, except for SN~2018zd, which is suggested to be an electron-capture SN (ECSN) by \cite{hiramatsu2021_2018zd}. The \ion{Fe}{ii} velocities of SN~2018zd fall below the mean, similar to the case of SN~2020aze, at least until 30 days. SN~2009au, classified as a LLEV SN, has the lowest \ion{Fe}{ii} and H$\alpha$ velocities among these objects.

\cite{Dessart2013} generated a grid of 15 M$_\odot$ stellar evolution models using Modules for Experiments in Stellar Astrophysics (MESA) STAR, exploring the effects of metallicity variations (0.1–2.0 Z$_\odot$) and parameters such as mass-loss rates, convection, and rotation on final stellar properties. \cite{Dessart2014} analysed these models and found that lower metallicity SNe progenitors exhibit weaker metal lines or reduced equivalent widths. Figure~\ref{metallicity} compares the 32-day spectrum of SN~2020aze with \cite{Dessart2014} models at metallicities of 0.1, 0.4, 1.0, and 2.0 Z$_\odot$ for the similar epoch. The \ion{Fe}{ii} $\lambda$5018 line EWs for these models, in increasing order of metallicity, are 4.4, 8.3, 10.9, and 11.9 \AA{} whereas the EW of the \ion{Fe}{ii} line in SN~2020aze is 11.25 \AA{}. This comparison suggests that SN~2020aze could have originated from a super-solar metallicity progenitor, as suggested by its relatively redder colours.

\section{Examining parallels to atypical Type II SNe}
\label{Parallels of SN 2020aze}

A comparison of SN~2020aze’s light curve and spectral parameters with those of other Type II SNe reveals notable overlaps with two distinct subclasses--ECSNe and LLEV SNe. We outline below the characteristic features of these atypical explosions and assess how the properties of SN~2020aze align with or diverge from them.

ECSNe are considered a subclass of SNe, expected to exhibit low explosion energies ($\sim$1.5~$\times$~10$^{50}$~erg), low progenitor masses (7--9.5~M$_\odot$), and low $^{56}$Ni yields ($\sim$2.5~$\times$~10$^{-3}$~M$_\odot$) \citep{Tominaga2013, Moriya2014}. Unlike iron CCSNe, which arise from the gravitational collapse of an Fe core, ECSNe are triggered when electron captures in an O–Ne–Mg core of a super asymptotic giant branch (AGB) star reduce electron pressure and induce core collapse. However, observational evidence remains scarce, with only a single confirmed candidate proposed to date (SN~2018zd, \citealt{hiramatsu2021_2018zd}). ECSNe exhibit low photospheric velocities (3000--4000 km s$^{-1}$) during the photospheric phase and often show interaction with CSM due to super AGB star winds. Observationally, ECSNe are identified by a bright, short-duration photospheric phase (60--100 days) followed by a faint tail with a significant magnitude drop. The photospheric phase duration may extend if the envelope mass (M$_{env}$) and hydrogen content are higher. SN~2020aze shares some ECSN-like traits, such as early-time CSM interaction indicated by flash features, and a relatively low photospheric velocity ($\sim$ 4200 km s$^{-1}$; \citealt{hiramatsu2021_2018zd}), suggesting a possible connection. However, other properties, such as its redder post-30-day colours, higher progenitor mass ($\sim$14 M$_\odot$) and higher explosion energy ($\sim$1.5$\times$10$^{51}$ erg) estimated from semi-analytical modelling, differ significantly from typical ECSNe. Further complicating the classification are uncertainties tied to distance-dependent modelling.

To mitigate such uncertainties, \citet{Sato2024} proposed a distance-independent method for identifying ECSNe based on the criterion:

\begin{equation}
(g-r)_{t_{pt}/2} < 0.008 \times t_{pt} - 0.4.  
\end{equation}
This criterion is motivated by light-curve simulations of ECSNe, which predict shorter, brighter, and bluer plateaus relative to those of Fe CCSNe \citep{Tominaga2013, Moriya2014}.\\

LLEV SNe are characterised by signatures of early CSM interaction, low expansion velocities, normal $^{56}$Ni masses (>0.01 M$_\odot$), bright $V$-band peak magnitude (<$-$17.5 mag), bluer \textit{B}-\textit{V} colours, and weaker metal lines \citep{Rodriguez2020}. Hydrodynamic simulations indicate that these SNe typically originate from progenitors of around 14 M$_\odot$ and exhibit explosion energies of 8$\times$10$^{50}$ ergs. SN~2009au, a LLEV SN, is a fast-declining SN with an unusually long photospheric phase duration, which is an anomaly based on the analysis done by \cite{anderson2014}. \cite{Hillier2019}, through hydrodynamical modelling, suggested that this behaviour could result from ejecta-CSM interaction. 

\section{Conclusions}
\label{discussion}

SN~2020aze is a fast-declining Type II SN with an unusually extended photospheric phase lasting 145 day. During this phase, SN~2020aze declines at rates of 4.91$\pm$0.47, 2.04$\pm$0.13, 1.33$\pm$0.11 mag (100 d)$^{-1}$ in {\it B}, {\it V}, and {\it r$^\prime$} bands, respectively. Its peak absolute $V$-band magnitude is $-$16.97 $\pm$ 0.20 mag, placing it at the upper limit of the normal Type II SNe range, which has a mean peak $V$-band magnitude of $-$16.74$\pm$1.01 mag \citep{anderson2014}.

The first two spectra of SN~2020aze, obtained 2.6 day and 3.0 day after explosion, provided an opportunity to capture early-time flash features arising from ejecta-CSM interaction. The spectra exhibited a narrow \ion{He}{ii} $\lambda$4686 line superimposed on a broad ledge feature having a FWHM velocity of $\sim$ 10,000 km s$^{-1}$. Such high velocity indicates blue-shifted \ion{He}{ii} emission, likely originating from the outer layers of the SN ejecta. By comparing the observed spectra with radiative transfer models from \cite{Dessart2023}, we inferred a wind mass-loss rate of $\sim$10$^{-3}$ M$_{\odot}$ yr$^{-1}$ for SN~2020aze. Light curve modelling further estimates the progenitor mass to be $\sim$14 M$_{\odot}$, suggesting an intermediate-mass progenitor.

The absorption dip in the P-Cygni of H$\alpha$ profile during the photospheric phase of SN~2020aze is less pronounced compared to normal Type IIP SNe, such as SNe~2017gmr and 2021yja, but more prominent than in fast-declining SNe like SN~2014G. A comparison of the 32 day spectrum of SN~2020aze with the \cite{Dessart2014} models suggests a progenitor with super-solar metallicity. While SN~2020aze exhibits certain properties consistent with ECSNe, such as low photospheric velocities and evidence of CSM interaction, the observed redder colours later than 30 days are inconsistent with the characteristics of ECSNe. In addition, the progenitor mass and explosion energy derived from semi-analytical light curve modelling of SN~2020aze are higher than those of ECSNe. SN~2020aze also shares some similarities with LLEV SNe, including CSM interaction and a bright peak magnitude. However, its higher expansion velocity, redder \textit{B}–\textit{V} colours, and stronger metal lines deviate from typical LLEV SN properties. While the evolution of ECSNe and LLEV SNe is largely dominated by CSM interaction, as indicated by their bluer colours, in SN~2020aze, this dominance is primarily restricted to the first 30 days. The steep decline combined with the extended photospheric phase observed in SN~2020aze is quite unusual within the Type II SN population, underscoring the wide diversity in light curve morphologies that can arise from the interplay of multiple factors, including pre-SN mass loss, which not only shapes the circumstellar environment but also influences the hydrogen envelope mass and density structure. Our results highlight how transitional events like SN~2020aze expand the known diversity of Type II SNe and offer valuable constraints for future models of massive star evolution.

\section*{Acknowledgements}
We thank the anonymous referee for providing us with valuable suggestions that improved the quality of the paper. This work uses data from the Las Cumbres Observatory Global Telescope network. The LCO group is supported by NSF grants AST-2308113 and AST-1911151. Time domain research by the University of Arizona team and D.J.S. is supported by National Science Foundation (NSF) grants 2108032, 2308181, 2407566, and 2432036 and the Heising-Simons Foundation under grant $\#$2020-1864. This research has made use of the APASS database, located on the AAVSO website. Funding for APASS has been provided by the Robert Martin Ayers Sciences Fund. Based on observations collected at the European Organisation for Astronomical Research in the Southern Hemisphere, Chile, as part of ePESSTO+ (the advanced Public ESO Spectroscopic Survey for Transient Objects Survey – PI: Inserra). ePESSTO+ observations were obtained under ESO program ID 1103.D-0328. S.V.  acknowledge support by NSF grant  AST-2008108. K.M. and B.A. acknowledge the support from the BRICS grant No. DST/ICD/BRICS/Call-5/CoNMuTraMO/2023 (G) funded by the Department of Science and Technology (DST), India. B.A. acknowledges the Council of Scientific $\&$ Industrial Research (CSIR) fellowship award (09/948(0005)/2020-EMR-I) for this work. The SALT observations presented here were made through the Rutgers University program 2019-1-MLT-004 (PI: S.W. Jha); SALT supernova spectroscopy at Rutgers is supported by NSF awards AST-1615455 and AST-2407567. KAB is supported by an LSST-DA Catalyst Fellowship; this publication was thus made possible through the support of Grant 62192 from the John Templeton Foundation to LSST-DA. T.E.M.B. is funded by Horizon Europe ERC grant no. 101125877. T.-W.C. acknowledges the financial support from the Yushan Fellow Program by the Ministry of Education, Taiwan (MOE-111-YSFMS-0008-001-P1) and the National Science and Technology Council, Taiwan (NSTC grant 114-2112-M-008-021-MY3).

\section*{Data Availability}
The photometric and spectroscopic data underlying this article will
be made available upon request to the corresponding author.



\bibliographystyle{mnras}
\bibliography{SN2020aze} 




\appendix

\section{Photometric and Spectroscopic Logs of SN~2020aze}
\begin{table*}
 \begin{center}
 \caption{DLT40 Clear band photometry of SN~2020aze.}
 \label{dlt_phot}
 \scalebox{0.95}{
 \begin{tabular}{|ccc|ccc|ccc|ccc|}
 \hline
 MJD  & Phase$^\dagger$ & Mag & MJD  & Phase$^\dagger$ & Mag & MJD  & Phase$^\dagger$ & Mag & MJD  & Phase$^\dagger$ & Mag\\
 (days) & (days) &  & (days) & (days) &  & (days) & (days) & & (days) & (days) & \\
\hline

58874.53 & 2.22 & 17.12$\pm$0.05 & 58876.69 & 4.38 & 16.89$\pm$0.03 & 58886.05 & 13.74 & 16.21$\pm$0.04 & 58918.50 & 46.19 & 16.64$\pm$0.03\\
58874.63 & 2.32 & 17.21$\pm$0.04 & 58876.83 & 4.52 & 16.74$\pm$0.03 & 58887.04 & 14.73 & 16.15$\pm$0.02 & 58919.05 & 46.74 & 16.88$\pm$0.03\\
58874.64 & 2.33 & 17.24$\pm$0.04 & 58876.84 & 4.53 & 16.57$\pm$0.03 & 58887.16 & 14.85 & 16.42$\pm$0.02 & 58919.49 & 47.18 & 16.70$\pm$0.03\\
58874.65 & 2.34 & 17.28$\pm$0.04 & 58877.53 & 5.22 & 16.31$\pm$0.04 & 58888.03 & 15.72 & 16.42$\pm$0.02 & 58920.01 & 47.70 & 16.72$\pm$0.03\\
58874.65 & 2.34 & 17.09$\pm$0.04 & 58877.59 & 5.28 & 16.49$\pm$0.03 & 58889.03 & 16.72 & 16.44$\pm$0.02 & 58920.49 & 48.18 & 16.67$\pm$0.03\\
58874.66 & 2.35 & 17.15$\pm$0.04 & 58878.06 & 5.75 & 16.51$\pm$0.03 & 58891.18 & 18.87 & 16.47$\pm$0.02 & 58921.01 & 48.70 & 16.72$\pm$0.03\\
58874.66 & 2.35 & 17.23$\pm$0.04 & 58878.06 & 5.75 & 16.44$\pm$0.02 & 58891.19 & 18.88 & 16.16$\pm$0.02 & 58922.01 & 49.70 & 16.74$\pm$0.03\\
58874.69 & 2.38 & 17.32$\pm$0.04 & 58878.52 & 6.21 & 16.47$\pm$0.04 & 58893.20 & 20.89 & 16.40$\pm$0.02 & 58923.00 & 50.69 & 16.80$\pm$0.03\\
58874.69 & 2.38 & 17.35$\pm$0.04 & 58878.52 & 6.21 & 16.53$\pm$0.04 & 58893.20 & 20.89 & 16.40$\pm$0.02 & 58923.52 & 51.21 & 16.58$\pm$0.03\\
58874.69 & 2.38 & 17.06$\pm$0.04 & 58878.53 & 6.22 & 16.43$\pm$0.03 & 58894.09 & 21.78 & 16.22$\pm$0.02 & 58924.00 & 51.69 & 16.72$\pm$0.03\\
58874.70 & 2.39 & 17.13$\pm$0.04 & 58878.53 & 6.22 & 16.32$\pm$0.04 & 58897.04 & 24.73 & 16.54$\pm$0.02 & 58925.00 & 52.69 & 16.74$\pm$0.03\\
58874.70 & 2.39 & 17.41$\pm$0.06 & 58878.83 & 6.52 & 16.44$\pm$0.02 & 58897.15 & 24.84 & 16.37$\pm$0.02 & 58929.64 & 57.33 & 16.89$\pm$0.03\\
58874.71 & 2.40 & 17.22$\pm$0.04 & 58879.11 & 6.80 & 16.35$\pm$0.02 & 58898.03 & 25.72 & 16.05$\pm$0.02 & 58930.48 & 58.17 & 16.89$\pm$0.03\\
58874.72 & 2.41 & 17.22$\pm$0.04 & 58879.52 & 7.21 & 16.45$\pm$0.03 & 58899.02 & 26.71 & 16.39$\pm$0.03 & 58931.48 & 59.17 & 16.81$\pm$0.03\\
58874.75 & 2.44 & 17.09$\pm$0.04 & 58879.59 & 7.28 & 16.45$\pm$0.03 & 58900.03 & 27.72 & 16.39$\pm$0.03 & 58932.48 & 60.17 & 16.82$\pm$0.03\\
58874.76 & 2.45 & 17.00$\pm$0.04 & 58879.74 & 7.43 & 16.44$\pm$0.02 & 58901.07 & 28.76 & 16.38$\pm$0.02 & 58933.53 & 61.22 & 16.96$\pm$0.03\\
58874.80 & 2.49 & 17.16$\pm$0.04 & 58880.12 & 7.81 & 16.30$\pm$0.02 & 58902.02 & 29.71 & 16.44$\pm$0.02 & 58934.48 & 62.17 & 16.94$\pm$0.03\\
58874.83 & 2.52 & 17.08$\pm$0.04 & 58880.12 & 7.81 & 16.37$\pm$0.02 & 58903.03 & 30.72 & 16.48$\pm$0.02 & 58935.47 & 63.16 & 16.88$\pm$0.03\\
58874.85 & 2.54 & 17.15$\pm$0.04 & 58881.05 & 8.74 & 16.15$\pm$0.02 & 58904.03 & 31.72 & 16.42$\pm$0.02 & 58937.47 & 65.16 & 17.06$\pm$0.04\\
58874.85 & 2.54 & 17.24$\pm$0.04 & 58881.19 & 8.88 & 16.36$\pm$0.02 & 58905.02 & 32.71 & 16.29$\pm$0.02 & 58955.52 & 83.21 & 17.28$\pm$0.04\\
58875.27 & 2.96 & 17.00$\pm$0.03 & 58881.52 & 9.21 & 16.32$\pm$0.04 & 58905.09 & 32.78 & 16.24$\pm$0.02 & 58958.45 & 86.14 & 17.07$\pm$0.04\\
58875.27 & 2.96 & 16.55$\pm$0.03 & 58881.59 & 9.28 & 16.27$\pm$0.02 & 58906.02 & 33.71 & 16.28$\pm$0.02 & 58960.45 & 88.14 & 17.27$\pm$0.04\\
58875.28 & 2.97 & 16.92$\pm$0.03 & 58881.77 & 9.46 & 16.39$\pm$0.02 & 58907.02 & 34.71 & 16.54$\pm$0.02 & 58965.52 & 93.21 & 17.40$\pm$0.05\\
58875.53 & 3.22 & 16.75$\pm$0.04 & 58882.10 & 9.79 & 16.23$\pm$0.02 & 58908.02 & 35.71 & 16.35$\pm$0.02 & 58966.46 & 94.15 & 17.42$\pm$0.05\\
58875.53 & 3.22 & 17.03$\pm$0.03 & 58882.14 & 9.83 & 16.31$\pm$0.02 & 58909.02 & 36.71 & 16.41$\pm$0.02 & 58967.46 & 95.15 & 17.51$\pm$0.05\\
58875.70 & 3.39 & 16.81$\pm$0.03 & 58882.19 & 9.88 & 16.33$\pm$0.02 & 58910.02 & 37.71 & 16.54$\pm$0.03 & 58969.45 & 97.14 & 17.52$\pm$0.05\\
58875.71 & 3.40 & 16.95$\pm$0.03 & 58882.52 & 10.21 & 16.39$\pm$0.02 & 58911.17 & 38.86 & 16.63$\pm$0.02 & 58970.46 & 98.15 & 17.52$\pm$0.05\\
58875.76 & 3.45 & 16.77$\pm$0.03 & 58882.52 & 10.21 & 16.49$\pm$0.02 & 58912.07 & 39.76 & 16.64$\pm$0.03 & 58980.43 & 108.12 & 17.63$\pm$0.06\\
58875.85 & 3.54 & 16.67$\pm$0.03 & 58883.06 & 10.75 & 16.24$\pm$0.02 & 58913.02 & 40.71 & 16.53$\pm$0.03 & 58981.43 & 109.12 & 17.76$\pm$0.06\\
58876.13 & 3.82 & 16.63$\pm$0.04 & 58883.17 & 10.86 & 16.32$\pm$0.02 & 58914.02 & 41.71 & 16.83$\pm$0.05 & 58984.66 & 112.35 & 17.91$\pm$0.07\\
58876.13 & 3.82 & 16.31$\pm$0.03 & 58883.52 & 11.21 & 16.31$\pm$0.03 & 58915.02 & 42.71 & 17.04$\pm$0.05 & 58985.43 & 113.12 & 17.68$\pm$0.06\\
58876.13 & 3.82 & 16.67$\pm$0.03 & 58884.08 & 11.77 & 16.24$\pm$0.02 & 58916.02 & 43.71 & 16.68$\pm$0.03 & 58986.66 & 114.35 & 17.97$\pm$0.07\\
58876.53 & 4.22 & 16.48$\pm$0.03 & 58884.27 & 11.96 & 16.25$\pm$0.02 & 58916.58 & 44.27 & 16.77$\pm$0.03 & 58987.43 & 115.12 & 18.03$\pm$0.07\\
58876.53 & 4.22 & 16.45$\pm$0.03 & 58885.05 & 12.74 & 15.98$\pm$0.03 & 58917.02 & 44.71 & 17.17$\pm$0.06 & 58990.43 & 118.12 & 17.99$\pm$0.07\\
58876.59 & 4.28 & 16.67$\pm$0.03 & 58885.24 & 12.93 & 16.34$\pm$0.03 & 58917.49 & 45.18 & 16.70$\pm$0.03 & 58991.45 & 119.14 & 17.89$\pm$0.06\\

 \hline
 \end{tabular}}
 \end{center}
 $^\dagger$Phase with respect to the explosion epoch (MJD = 58872.31).
\end{table*}
\begin{table*}
 \begin{center}
 \caption{LCO $UBVg^\prime r^\prime i^\prime$ bands photometry of SN~2020aze.}
 \label{optical_phot}
 \scalebox{0.85}{
 \begin{tabular}{@{}lcccccccc}
 \hline
 MJD  & Phase$^\dagger$ & $U$ & $B$ & $V$ & $g^\prime$ & $r^\prime$ & $i^\prime$ & Telescope$^\ddagger$\\
 (days) & (days) & (mag) & (mag) & (mag) & (mag) & (mag) & (mag)\\ 
 \hline
58874.92 & 2.61 & 16.70$\pm$0.02 & 17.47$\pm$0.03 & 17.32$\pm$0.01 & 17.31$\pm$0.03 & 17.24$\pm$0.02 & 17.20$\pm$0.02 & CPT 1m\\
58875.55 & 3.24 & 16.26$\pm$0.02 & -- & -- & -- & -- & -- & COJ 1m\\
58876.09 & 3.78 & 16.39$\pm$0.01 & 17.27$\pm$0.01 & 17.10$\pm$0.05 & 17.04$\pm$0.01 & 16.98$\pm$0.01 & 16.93$\pm$0.01 & CPT 1m\\
58876.92 & 4.61 & 16.25$\pm$0.02 & 17.23$\pm$0.01 & 16.91$\pm$0.02 & 16.89$\pm$0.01 & 16.89$\pm$0.01 & 16.72$\pm$0.01 & CPT 1m\\
58877.95 & 5.64 & 16.23$\pm$0.01 & 16.99$\pm$0.02 & 16.92$\pm$0.02 & 16.75$\pm$0.02 & 16.70$\pm$0.01 & -- & CPT 1m\\
58878.06 & 5.75 & -- & -- & -- & -- & -- & 16.67$\pm$0.01 & CPT 1m\\
58879.74 & 7.43 & 16.14$\pm$0.02 & 16.95$\pm$0.02 & 16.71$\pm$0.01 & 16.62$\pm$0.01 & 16.59$\pm$0.01 & 16.38$\pm$0.01 & COJ 1m\\
58881.12 & 8.81 & 16.13$\pm$0.02 & 16.94$\pm$0.02 & 16.81$\pm$0.01 & 16.58$\pm$0.01 & 16.55$\pm$0.01 & 16.51$\pm$0.01 & LSC 1m\\
58883.00 & 10.69 & 16.34$\pm$0.01 & 16.96$\pm$0.01 & 16.72$\pm$0.01 & 16.57$\pm$0.02 & 16.42$\pm$0.01 & 16.35$\pm$0.02 & CPT 1m\\
58884.90 & 12.59 & 16.34$\pm$0.02 & 16.93$\pm$0.02 & 16.70$\pm$0.02 & 16.67$\pm$0.03 & 16.26$\pm$0.01 & 16.54$\pm$0.03 & CPT 1m\\
58889.08 & 16.77 & 16.47$\pm$0.01 & 17.17$\pm$0.02 & 16.81$\pm$0.01 & 16.66$\pm$0.01 & -- & -- & CPT 1m\\
58893.12 & 20.81 & 16.64$\pm$0.02 & 17.18$\pm$0.01 & 17.03$\pm$0.02 & 16.87$\pm$0.01 & 16.62$\pm$0.01 & 16.42$\pm$0.03 & CPT 1m\\
58903.08 & 30.77 & 17.82$\pm$0.02 & 17.76$\pm$0.02 & 17.32$\pm$0.02 & 16.82$\pm$0.02 & 16.35$\pm$0.01 & -- & CPT 1m\\
58907.08 & 34.77 & 18.30$\pm$0.03 & 17.85$\pm$0.01 & 17.53$\pm$0.01 & 16.91$\pm$0.01 & 16.66$\pm$0.01 & 16.39$\pm$0.01 & LSC 1m\\
58910.85 & 38.54 & 18.54$\pm$0.04 & 18.16$\pm$0.04 & 17.59$\pm$0.02 & 17.20$\pm$0.04 & 16.51$\pm$0.03 & 16.09$\pm$0.03 & CPT 1m\\
58915.29 & 42.98 & 18.95$\pm$0.06 & 18.41$\pm$0.03 & 17.94$\pm$0.02 & 17.12$\pm$0.01 & 16.71$\pm$0.01 & 16.46$\pm$0.01 & LSC 1m\\
58919.21 & 46.90 & 19.42$\pm$0.09 & 19.06$\pm$0.02 & -- & 17.34$\pm$0.01 & -- & -- & LSC 1m\\
58922.24 & 49.93 & -- & -- & 18.09$\pm$0.01 & 17.55$\pm$0.01 & 16.85$\pm$0.01 & 16.59$\pm$0.01 & LSC 1m\\
58928.50 & 56.19 & -- & 18.96$\pm$0.02 & 18.37$\pm$0.01 & 17.39$\pm$0.01 & 16.98$\pm$0.01 & 16.58$\pm$0.01 & COJ 1m\\
58938.67 & 66.36 & -- & 19.19$\pm$0.03 & 18.46$\pm$0.01 & 17.55$\pm$0.02 & 16.94$\pm$0.01 & 16.78$\pm$0.01 & COJ 1m\\
58944.63 & 72.32 & -- & -- & -- & -- & -- & 16.75$\pm$0.04 & COJ 1m\\
58952.39 & 80.08 & -- & 19.80$\pm$0.04 & 18.85$\pm$0.01 & 17.94$\pm$0.01 & 17.27$\pm$0.01 & 16.96$\pm$0.01 & COJ 1m\\
58957.47 & 85.16 & -- & 20.09$\pm$0.16 & 19.09$\pm$0.01 & 17.96$\pm$0.02 & 17.34$\pm$0.02 & -- & COJ 1m\\
58961.59 & 89.28 & -- & -- & 19.25$\pm$0.03 & 18.02$\pm$0.02 & 17.32$\pm$0.01 & 17.00$\pm$0.01 & COJ 1m\\
58972.55 & 100.24 & -- & -- & 19.53$\pm$0.11 & 18.71$\pm$0.12 & 17.26$\pm$0.03 & 17.14$\pm$0.02 & COJ 1m\\
58978.51 & 106.20 & -- & -- & -- & 18.49$\pm$0.08 & 17.58$\pm$0.05 & 17.26$\pm$0.04 & COJ 1m\\
58984.79 & 112.48 & -- & -- & -- & 18.92$\pm$0.02 & 17.70$\pm$0.04 & 17.67$\pm$0.01 & CPT 1m\\
58989.87 & 117.56 & -- & -- & -- & 18.84$\pm$0.03 & 17.93$\pm$0.01 & 17.58$\pm$0.02 & CPT 1m\\
58993.36 & 121.05 & -- & -- & 20.07$\pm$0.06 & -- & -- & -- & COJ 1m\\
58993.49 & 121.18 & -- & -- & -- & 18.85$\pm$0.03 & -- & -- & COJ 1m\\
58996.52 & 124.21 & -- & -- & -- & -- & 18.23$\pm$0.12 & 18.21$\pm$0.12 & COJ 1m\\
58999.48 & 127.17 & -- & -- & 20.60$\pm$0.06 & 19.24$\pm$0.03 & 18.21$\pm$0.01 & 17.90$\pm$0.02 & COJ 1m\\
59001.82 & 129.51 & -- & -- & -- & 19.35$\pm$0.05 & 18.37$\pm$0.01 & 17.91$\pm$0.01 & CPT 1m\\
59004.76 & 132.45 & -- & -- & 21.39$\pm$0.07 & 19.67$\pm$0.04 & 18.52$\pm$0.02 & 18.25$\pm$0.02 & CPT 1m\\
59009.70 & 137.39 & -- & -- & 21.25$\pm$0.11 & 20.20$\pm$0.07 & 19.29$\pm$0.02 & 18.86$\pm$0.02 & CPT 1m\\

\hline
 \end{tabular}}
 
 $^\dagger$Phase with respect to the explosion epoch (MJD = 58872.31).
 \newline
 $^\ddagger$CPT stands for South African Astronomical Observatory, COJ stands for Siding Spring Observatory, and LSC stands for Cerro Tololo Interamerican Observatory.
\end{center}
\end{table*}
\begin{table*}
 \begin{center}
 \caption{ATLAS $o$-band photometry of SN~2020aze.}
 \label{atlas_phot}
 \scalebox{0.85}{
 \begin{tabular}{cccccccccccc}
 \hline

 MJD  & Phase$^\dagger$ & Mag & MJD  & Phase$^\dagger$ & Mag & MJD  & Phase$^\dagger$ & Mag & MJD  & Phase$^\dagger$ & Mag\\
 (days) & (days) &  & (days) & (days) &  & (days) & (days) & & (days) & (days) & \\
\hline

58875.48 & 3.17 & 17.06$\pm$0.04 & 58903.38 & 31.07 & 16.50$\pm$0.02 & 58939.36 & 67.05 & 17.06$\pm$0.14 & 58975.30 & 102.99 & 17.40$\pm$0.07\\
58875.49 & 3.18 & 17.06$\pm$0.03 & 58903.38 & 31.07 & 16.45$\pm$0.02 & 58941.37 & 69.06 & 16.93$\pm$0.03 & 58975.31 & 103.00 & 17.61$\pm$0.1\\
58875.50 & 3.19 & 16.99$\pm$0.03 & 58903.39 & 31.08 & 16.58$\pm$0.02 & 58941.38 & 69.07 & 16.85$\pm$0.03 & 58975.32 & 103.01 & 17.60$\pm$0.09\\
58875.51 & 3.20 & 17.07$\pm$0.03 & 58903.40 & 31.09 & 16.26$\pm$0.02 & 58941.38 & 69.07 & 16.90$\pm$0.04 & 58979.29 & 106.98 & 17.57$\pm$0.04\\
58879.45 & 7.14 & 16.50$\pm$0.02 & 58907.35 & 35.04 & 16.53$\pm$0.02 & 58941.40 & 69.09 & 16.90$\pm$0.04 & 58979.30 & 106.99 & 17.54$\pm$0.06\\
58879.46 & 7.15 & 16.46$\pm$0.02 & 58907.36 & 35.05 & 16.47$\pm$0.02 & 58947.32 & 75.01 & 16.90$\pm$0.12 & 58979.30 & 106.99 & 17.52$\pm$0.05\\
58879.47 & 7.16 & 16.51$\pm$0.02 & 58907.37 & 35.06 & 16.45$\pm$0.02 & 58951.25 & 78.94 & 17.70$\pm$0.21 & 58979.32 & 107.01 & 17.45$\pm$0.1\\
58879.47 & 7.16 & 16.55$\pm$0.02 & 58907.38 & 35.07 & 16.51$\pm$0.02 & 58951.27 & 78.96 & 16.98$\pm$0.03 & 58983.27 & 110.96 & 17.63$\pm$0.05\\
58883.41 & 11.10 & 16.58$\pm$0.04 & 58913.41 & 41.10 & 16.49$\pm$0.03 & 58951.28 & 78.97 & 17.08$\pm$0.03 & 58983.28 & 110.97 & 17.63$\pm$0.05\\
58883.42 & 11.11 & 16.57$\pm$0.04 & 58913.42 & 41.11 & 16.51$\pm$0.03 & 58951.28 & 78.97 & 17.08$\pm$0.03 & 58983.28 & 110.97 & 17.59$\pm$0.05\\
58883.42 & 11.11 & 16.84$\pm$0.06 & 58913.42 & 41.11 & 16.56$\pm$0.03 & 58955.25 & 82.94 & 17.16$\pm$0.05 & 58983.29 & 110.98 & 17.75$\pm$0.07\\
58883.43 & 11.12 & 16.55$\pm$0.04 & 58913.45 & 41.14 & 16.54$\pm$0.04 & 58955.26 & 82.95 & 17.21$\pm$0.05 & 58987.27 & 114.96 & 17.85$\pm$0.07\\
58883.45 & 11.14 & 16.42$\pm$0.02 & 58915.33 & 43.02 & 16.65$\pm$0.06 & 58955.27 & 82.96 & 17.16$\pm$0.05 & 58987.27 & 114.96 & 17.83$\pm$0.06\\
58883.48 & 11.17 & 16.45$\pm$0.03 & 58915.34 & 43.03 & 16.70$\pm$0.09 & 58955.28 & 82.97 & 17.18$\pm$0.04 & 58987.28 & 114.97 & 17.89$\pm$0.07\\
58883.48 & 11.17 & 16.42$\pm$0.02 & 58915.35 & 43.04 & 16.87$\pm$0.32 & 58959.24 & 86.93 & 17.21$\pm$0.05 & 58987.29 & 114.98 & 17.88$\pm$0.07\\
58887.41 & 15.10 & 16.53$\pm$0.07 & 58915.35 & 43.04 & 16.07$\pm$0.15 & 58959.24 & 86.93 & 17.22$\pm$0.04 & 58991.27 & 118.96 & 17.89$\pm$0.06\\
58887.43 & 15.12 & 16.54$\pm$0.06 & 58919.32 & 47.01 & 16.66$\pm$0.05 & 58959.25 & 86.94 & 17.05$\pm$0.04 & 58991.27 & 118.96 & 17.83$\pm$0.06\\
58887.45 & 15.14 & 16.32$\pm$0.05 & 58919.33 & 47.02 & 16.56$\pm$0.04 & 58959.27 & 86.96 & 17.14$\pm$0.04 & 58991.28 & 118.97 & 17.99$\pm$0.08\\
58887.47 & 15.16 & 16.41$\pm$0.04 & 58919.35 & 47.04 & 16.55$\pm$0.04 & 58963.31 & 91.00 & 16.97$\pm$0.38 & 58991.29 & 118.98 & 17.91$\pm$0.07\\
58887.48 & 15.17 & 16.35$\pm$0.04 & 58919.36 & 47.05 & 16.62$\pm$0.04 & 58967.29 & 94.98 & 17.33$\pm$0.04 & 58995.25 & 122.94 & 17.96$\pm$0.08\\
58887.48 & 15.17 & 16.47$\pm$0.04 & 58927.44 & 55.13 & 16.84$\pm$0.05 & 58967.30 & 94.99 & 17.35$\pm$0.04 & 58995.25 & 122.94 & 18.11$\pm$0.10\\
58887.49 & 15.18 & 16.40$\pm$0.04 & 58927.44 & 55.13 & 16.92$\pm$0.06 & 58967.30 & 94.99 & 17.34$\pm$0.05 & 58995.26 & 122.95 & 17.99$\pm$0.12\\
58891.39 & 19.08 & 16.35$\pm$0.04 & 58927.45 & 55.14 & 16.95$\pm$0.04 & 58967.31 & 95.00 & 17.37$\pm$0.04 & 58999.25 & 126.94 & 17.97$\pm$0.12\\
58891.39 & 19.08 & 16.38$\pm$0.04 & 58927.46 & 55.15 & 16.92$\pm$0.05 & 58971.28 & 98.97 & 17.29$\pm$0.06 & 58999.25 & 126.94 & 18.24$\pm$0.12\\
58891.40 & 19.09 & 16.32$\pm$0.03 & 58927.47 & 55.16 & 16.83$\pm$0.03 & 58971.29 & 98.98 & 17.24$\pm$0.04 & 58999.26 & 126.95 & 18.02$\pm$0.08\\
58891.41 & 19.10 & 16.39$\pm$0.04 & 58927.47 & 55.16 & 16.82$\pm$0.03 & 58971.29 & 98.98 & 17.20$\pm$0.04 & 58999.27 & 126.96 & 18.03$\pm$0.06\\
58895.38 & 23.07 & 16.34$\pm$0.02 & 58927.48 & 55.17 & 16.77$\pm$0.03 & 58971.31 & 99.00 & 17.41$\pm$0.06 & 59003.27 & 130.96 & 18.41$\pm$0.20\\
58895.39 & 23.08 & 16.40$\pm$0.02 & 58927.49 & 55.18 & 16.74$\pm$0.03 & 58973.28 & 100.97 & 17.34$\pm$0.11 & 59007.24 & 134.93 & 18.98$\pm$1.25\\
58895.40 & 23.09 & 16.54$\pm$0.02 & 58939.27 & 66.96 & 16.82$\pm$0.47 & 58973.28 & 100.97 & 17.48$\pm$0.08 & 59007.25 & 134.94 & 18.42$\pm$0.42\\
58895.42 & 23.11 & 16.44$\pm$0.02 & 58939.29 & 66.98 & 16.98$\pm$0.06 & 58973.29 & 100.98 & 17.50$\pm$0.08 & -- & -- & --\\
58899.37 & 27.06 & 16.56$\pm$0.04 & 58939.34 & 67.03 & 16.70$\pm$0.11 & 58973.32 & 101.01 & 17.42$\pm$0.09 & -- & -- & --\\
58899.38 & 27.07 & 16.48$\pm$0.03 & 58939.35 & 67.04 & 17.07$\pm$0.26 & 58975.30 & 102.99 & 17.26$\pm$0.06 & -- & -- & --\\

 \hline
 \end{tabular}}
 \end{center}
 $^\dagger$Phase with respect to the explosion epoch (MJD = 58872.31).
\end{table*}
\begin{table*}
 \begin{center}
 \caption{UVOT lightcurve of SN~2020aze.}
 \label{UVOT_phot}
 \scalebox{0.85}{
 \begin{tabular}{@{}lcccc}
 \hline
 MJD  & Phase$^\dagger$ & $UVW2$ & $UVM2$ & $UVW1$ \\
 (days) & (days) & (mag) & (mag) & (mag)  \\ 
 \hline
58875.91 & 3.60 & 16.33$\pm$0.12 & 16.47$\pm$0.13 & 16.47$\pm$0.13\\
58876.68 & 4.37 & 16.64$\pm$0.16 & 16.46$\pm$0.15 & 16.46$\pm$0.15\\
58877.85 & 5.54 & 16.57$\pm$0.15 & 16.54$\pm$0.15 & 16.54$\pm$0.15\\
58878.86 & 6.55 & 16.85$\pm$0.16 & 16.99$\pm$0.19 & 16.99$\pm$0.19\\
58882.88 & 10.57 & 17.09$\pm$0.21 & 17.13$\pm$0.23 & 17.13$\pm$0.23\\

 \hline
 \end{tabular}}
\end{center}
$^\dagger$Phase with respect to the explosion epoch (MJD = 58872.31).
\end{table*}
\begin{table*}
 \begin{center}
 \caption{Log of Spectroscopic Observations}
 \label{spec_log}
 \scalebox{0.8}{
 \begin{tabular}{@{}lcccc}
 \hline
 MJD & Phase$^\dagger$ &Telescope & Instrument & Slit Width\\
 (days) & (days) & & & (arc-sec)\\
 \hline
 58874.9 & 2.6 & SALT & RSS & 1.5  \\
 58875.3 & 3.0 & GS & GMOS-S & 1.5\\
 58880.0 & 7.7 & NTT  & EFOSC2 & 1.0\\
 58881.2 & 8.9 & NTT  & EFOSC2 & 1.0\\
 58881.5 & 9.2 & FTS & FLOYDS & 2.0\\
 58882.0 & 9.7 & NTT  & EFOSC2 & 1.0\\
 58890.0 & 17.7 & SOAR & Goodman HTS & 1.0\\
 58891.0 & 18.7 & NTT  & EFOSC2 & 1.0\\
 58893.6 & 21.3 & FTS & FLOYDS & 2.0\\
 58904.0 & 31.7 & NTT  & EFOSC2 & 1.0\\
 58908.6 & 36.3 & FTS & FLOYDS & 2.0\\
 58925.4 & 53.1 & FTS & FLOYDS & 2.0\\
 59194.3$^\dagger$$^\dagger$ & 322.0 & VLT & FORS2 & 1.6\\
 \hline
 \end{tabular}}
 \end{center}
  $^\dagger$Phase with respect to the explosion epoch (MJD = 58872.31).\\
  $^\dagger$$^\dagger$Downloaded from the ESO public archive.
\end{table*}


\bsp	
\label{lastpage}
\end{document}